\documentclass[traditabstract]{aa}
\usepackage[varg]{txfonts}

\usepackage{epsfig,graphicx,lscape,subfigure}
\usepackage{float,longtable,comment}
\usepackage{rotating,amssymb,amsmath}
\usepackage{natbib}

\newcommand{\kms}{{km\,s}$^{-1}$}
\newcommand{\cms}{{cm\,s}$^{-2}$}

\newcommand{\teff}{$T_\mathrm{eff}$\,}
\newcommand{\logg}{$\log g$\,}

\newcommand{\Msun}{\,$\rm{M}_\odot$}
\newcommand{\Lsun}{\,$\rm{L}_\odot$}

\newcommand{\Mi}{\,$\rm{M}_{\rm{i}}$}

\newcommand{\MgtoH}{$\epsilon_{\rm{Mg}}$}
\newcommand{\CtoH}{$\epsilon_{\rm{C}}$}
\newcommand{\NtoH}{$\epsilon_{\rm{N}}$}
\newcommand{\NtoHL}{$\epsilon_{\rm{0}}$}
\newcommand{\ltf}{$\leq$40}

\newcommand{\vsini}{$v_{\rm{e}} \sin i$}
\newcommand{\ve}{$v_{\rm{e}}$}
\newcommand{\vl}{$v_{\rm{0}}$}
\newcommand{\vt}{$v_{\rm{t}}$}
\newcommand{\vi}{$v_{\rm{i}}$}
\newcommand{\vr}{$v_{\rm{r}}$}

\newcommand{\pmg}{mas\,yr$^{-1}$}

\begin{document}
\title{The NGC\,346 massive star census}
\subtitle{Nitrogen abundances for apparently single, narrow lined,\\hydrogen core burning B-type stars\thanks{Based on observations obtained at the European Southern Observatory in programmes 074.D-0011, 081.D-0364 and 171.D-0237}}

\author{P.L. Dufton\inst{\ref{i1}}, C. J. Evans\inst{\ref{i3}}, D. J. Lennon\inst{\ref{i7},\ref{i8}}, I. Hunter\inst{\ref{i1}}}
\institute{Astrophysics Research Centre, School of Mathematics and Physics, Queen's University Belfast, Belfast BT7 1NN, UK\label{i1}
	\and UK Astronomy Technology Centre, Royal Observatory Edinburgh, Blackford Hill, Edinburgh, EH9 3HJ, UK\label{i3}
   \and Instituto de Astrof\'isica de Canarias, E-38200 La Laguna, Tenerife, Spain\label{i7}
   \and Departamento de Astrof\'isica, Universidad de La Laguna, E-38205 La Laguna, Tenerife, Spain\label{i8}
}
\date{Received / accepted }

\abstract{Previous analyses of two large spectroscopic surveys of early-type stars in the Large Magellanic Cloud (LMC) have found an excess of nitrogen enriched B-type targets with a \vsini\,$\leq$\,40 \kms\ compared with the predictions of single star evolutionary models incorporating rotational mixing. By contrast the number of such targets with 40\,$<$\,\vsini$\,\leq$\,80~\kms\ were consistent with such models. We have undertaken a similar analysis for 61 B-type targets which lie towards the young cluster, NGC\,346 in the Small Magellanic Cloud (SMC). These again have projected rotational velocities, \vsini\,$\leq$\,80 \kms, are not classified as supergiants and are apparently single.
Approximately 65\% of these SMC targets could have nitrogen enhancements of less than 0.3 dex, consistent with them having experienced only small amounts of mixing due to their low rotational velocities. However as with the previous LMC surveys, an excess of stars with low projected rotational velocities, \vsini\,$\leq$\,40 \kms\ and significant nitrogen enrichments is found. This is estimated to be approximately 5\% of the total population of apparently single B-type stars or 40\% of all stars with current rotational velocities of less than 40\,\kms; these percentages are similar to those found previously for the two LMC samples. For all three surveys, the presence of undetected binaries and other uncertainties imply that these percentages might be underestimated and indeed it is possible that all single stars with current rotational velocities of less than 40\,\kms\ are nitrogen enriched.

Two possible explanations incorporate the effects of magnetic field, via either a stellar merger followed by magnetic breaking or the evolution of a single star with a large magnetic field. Both mechanisms would appear to be compatible with the observed frequency of nitrogen-enriched stars in the Magellanic Clouds. Differences in the properties of the nitrogen-enriched stars compared with the remainder of the sample would be consistent with the former mechanism. For the latter, a qualitative comparison with {\em Galactic} evolutionary models incorporating magnetic fields is encouraging in terms of the amount of nitrogen enrichment and its presence in stars near the zero-age main sequence.
}
\keywords{stars: early-type -- stars: rotation -- stars: abundances -- Magellanic Clouds -- galaxies: star cluster: individual: NGC\,346}

\authorrunning{P.L. Dufton et al.}
\titlerunning{B-type stellar nitrogen abundances in NGC\,346}

\maketitle
\nopagebreak[4]

\section{Introduction}                                         \label{s_intro}

Spectroscopy of early-type stars has been quantitatively analysed for over seventy years \citep{uns42}, whilst their evolution from the main sequence has been modelled for over sixty years \citep[e.g.][]{tay54, tay56, kus57, sch58}. Subsequently there have been major advances in the physical assumptions adopted in such models, including the effects of stellar rotation \citep{mae87}, mass loss \citep{chi86, pau86, pul96} and magnetic fields \citep{don09, pet15, kes19}. Reviews of these developments have been provided by \citet{mae09} and \citet{lan12}.

Early spectroscopic analyses normally considered either a single or small number of targets both in our Galaxy \citep[see, for example][]{har70, kod70} and subsequently in the Magellanic Clouds \citep[e.g.][]{ kor00, kor02, kor05, rol02, rol03, len03}. However for the latter, the availability of multi-object spectroscopic facilities has allowed large scale surveys to be undertaken. For example, programmes with the European Southern Observatory (ESO) Very Large Telescope (VLT) have been used to investigate the rotational properties of early-type (both apparently single and binary) stars in the Magellanic Clouds \citep{mar06, mart07, hun08a, duf12, ram13, ram15, gar17}.

The stellar parameters estimated from these spectroscopic surveys can also be compared to grids of evolutionary models \citep{bro11a, geo13, geo13a}. \citet{hun08b, hun09a} found that although current evolutionary models were consistent with the properties of most of the B-type stars in the Large Magellanic Cloud (LMC) observed by \citet{eva06}, there were discrepancies. These included a group of stars with enhanced nitrogen abundances and small projected rotational velocities (\vsini), which \citet{hun08b} identified as `Group 2'. Their nature has subsequently been discussed by, for example, \citet{bro11b}, \citet{mae14}, and \citet{aer14a}, whilst similar targets have been identified in the O-type population in the Clouds \citep{riv12, gri17}

\citet[][hereafter Paper I]{duf18} reconsidered the LMC targets identified by \citet{hun08b,hun09a} and also analysed spectroscopy for B-type targets from the VLT-FLAMES Tarantula Survey of the 30~Doradus complex \citep[hereafter VFTS]{eva11}. They compared their results with computer simulations of the number of rapidly rotating nitrogen enriched stars that would be observed at small angles of inclination. They found that the number of observed targets with 40$<$\vsini$\leq$80\,\kms\ was consistent with them being rapidly rotating. By contrast for both datasets, they found an excess in the number of targets with \vsini$\leq$40\,\kms\ compared with their simulations. They discussed possible explanations for this excess of which the two most likely involved magnetic fields.

Here we analyse apparently single B-type stars with low projected rotational velocities towards the young cluster NGC\,346 in the Small Magellanic Cloud (SMC). Using methods similar to those in Paper I, we again investigate whether there is an excess of nitrogen enhanced stars with low projected velocities and attempt to quantify any such excess. We then use the results from all three surveys to discuss the evolutionary pathways that may have produced these stars.
\section{Observations}\label{s_obs}
Most of the observational data has been taken from two surveys \citep{eva06, eva11} that used the Fibre Large Array Multi-Element Spectrograph \citep[FLAMES,][]{pas02} on the VLT; the Medusa mode fed light to the Giraffe spectrograph for typically 80 stellar targets (plus sky fibres) in a single exposure. Additional HR02 spectroscopy has been subsequently obtained for some targets and is discussed separately below. Initial reductions utilised the Giraffe Base-line Data Reduction Software \citep[girBLDRS;][]{ble03} for bias subtraction, flat-field correction, fibre extraction and wavelength calibration. Subsequently sky subtraction and correction to the heliocentric velocity frame was undertaken with the {\sc starlink} software {\sc dipso} \citep{cur13}.

The first source of data was the FLAMES Survey of Massive Stars \citep[][hereafter FSMS]{eva05, eva06}. Spectroscopy was obtained for 77 OB-type stars in NGC\,346 using five high-resolution grating settings (HR02--HR06) to cover the wavelengths region from 3854--4760\AA\ with a spectral resolving power, $R$, of approximately 20\,000. Further details of the observing strategy, together with spectral types and binary identifications can be found in \citet{eva06}. \citet{hun08a} subsequently estimated projected rotational velocities, \vsini, for the B-type targets using a profile fitting technique. 

Our selection criteria were to identify apparently single B-type stars with \vsini$\leq$80\,\kms\ that were in the core hydrogen burning phase. \citet{hun08b} inferred a surface gravity, \logg$\simeq$3.2\,dex,  for when this evolutionary phase ends and we have therefore limited our sample to targets with estimated gravities, \logg$>$ 3.2\,dex (see Sect.\ \ref{s_atm}). These criteria led to 17 FSMS targets which are summarised in Table \ref{t_Targets}, with their numbering taken from \citet{eva06}. We have re-combined the HR02 observations (which contain the \ion{N}{ii} feature at 3995\AA) using a median or weighted $\sigma$-clipping algorithm. The final spectra from the two different methods were usually indistinguishable and were normalised using a single, low-order polynomial. As discussed in Paper I, these spectra should provide a better data product than the simple co-addition of exposures that was undertaken in the original analyses. Comparison of spectra from the two reductions confirmed this (although in many cases the improvement was small). Also listed in Table \ref{t_Targets} are projected rotational velocities, \vsini\ \citep{hun08a, 346_lr} and distances from the cluster centre \citep{eva06}. Signal-to-noise (S/N) ratios obtained for the region near to the \ion{N}{ii} line at 3995\AA\ were in the range 55--105 with a median value of 78.

The second source of data was the massive star census (hereafter designated as MSC) of NGC\,346 \citep[][hereafter Paper II]{346_lr}, with a magnitude limit of $V$\,$\le$\,16.75, corresponding to a latest main-sequence spectral type of B3~V. Two hundred and thirty OB-type stars were observed using the LR02 (wavelength range from 3960 to 4564\,\AA\ at $R$\,$\sim$\,7\,000) and LR03 (4499-5071\,\AA, $R$\,$\sim$\,8\,500) grating settings. Further details of target selection and observing strategy, together with spectral types and binary identifications can be found in Paper II. Applying the same selection criteria produced 45 targets which have been summarised in Table \ref{t_Targets} with their numbering, spectral types, projected rotational velocity estimates and distances from the cluster centre being taken from Paper II. The S/N ratios were estimated as for the FSMS data and were in the range 50--130 with a median value of 70. The moderate spectral resolution of the LR02 and LR03 settings led to estimates of the projected rotational velocity for stars with \vsini\,$\leq$ 40 \kms\ being poorly constrained and hence approximately half of our sample have been assigned to a bin with $0\le$\,\vsini\,$\le40$\ \kms.

Additional HR02 spectroscopy was available for 18 targets and has been reduced as described above. For the three FSMS targets (\#0043, \#0044, \#0102), these exposures were simply combined with the existing HR02 spectroscopy. For the remaining MSC targets, the HR02 spectroscopy was reduced independently and had S/N ratios ranging from 30--135 with a median of 55. As the HR02 was generally of higher quality, it has been used to estimate nitrogen abundances in preference to the LR02 data.

Luminosity estimates for all targets have been also taken from Paper II and are listed in Table \ref{t_Targets}. A uniform reddening of $E(B-V)$=0.09 \citep{mas95} with a reddening law of  $A_{\rm V}$\,=\,2.72$E(B-V)$  \citep{bou85}, a distance modulus of 18.91\,dex \citep{hil05} and bolometric corrections from \citet{vac96} and \citet{bal94} were adopted. In Paper II, a typical uncertainty of $\pm$0.2 dex was estimated for these values. 
 
In the model atmosphere analysis, equivalent widths were generally used for the metal lines. These were estimated by fitting Gaussian profiles (and a low order polynomial to represent the continuum) to the observed spectra leading to formal errors of typically 10\%. The fits were generally convincing which is consistent with the intrinsic and instrumental broadening being major contributors for these narrow-lined spectra. Tests using rotationally-broadened profiles yielded equivalent width estimates in good agreement with those using Gaussian profiles -- differences were normally less than 10\% for well observed lines but the fits were generally less convincing.

For the HR02 spectroscopy, the \ion{N}{ii} line at 3995\AA\ was normally well observed but for the LR02 data, this was not always the case. In these cases, we have set upper limits on the equivalent width using the methodology discussed in Paper I. This measured the equivalent widths of the weakest metal lines that that were observable for LR02 spectra with different S/N ratios. These were then used to set an upper limit for the \ion{N}{ii} line strength\footnote{When the \ion{N}{ii} line was not seen in either the LR02 or HR02 spectroscopy, the latter was mapped to the LR02 wavelength scale and combined when estimating the S/N ratios.}. As discussed in Paper I, this will lead to conservative estimates for both the upper limits of the equivalent widths and  the corresponding nitrogen abundances.

Table \ref{t_Targets} also provides kinematic data for all the targets. Radial-velocity estimates for the FSMS targets were taken directly from \citet{eva06}, who estimated a typical uncertainty of 5-10\,\kms. Those for the MSC targets are from line centre estimates obtained by fitting Gaussian profiles as discussed above and had a typical standard deviation of 4-8\,\kms\ with a maximum uncertainty of 12\,\kms\ for \#1079, which had a relatively weak metal line spectrum. Also listed in Table \ref{t_Targets} are the proper motion estimates from the {\em Gaia} DR2 data release \citep{gaia1, gaia2}.

\begin{table*}
\caption{B-type targets towards NGC\,346, which show no evidence of significant radial-velocity variations \citep{eva06, 346_lr}, have \vsini\,$\le$\,80 \kms\ \citep{hun08b, 346_lr} and a surface gravity estimate, \logg$>$3.2\,dex. Further details on the targets including their co-ordinates can be found in \citet{eva06} and \citet{346_lr}. Targets marked with an asterisk had additional HR02 spectroscopy, whilst luminosities, L, were estimated as discussed in Sect. \ref{s_obs}.\label{t_Targets}}
\begin{center}
\begin{tabular}{llrrccccc} 
\hline\hline
Star   & Classification & \vsini     &  $r$    & $\log$\,L/\Lsun & \vr     & PM(RA) & PM(Dec)   \\ 
       &                & (\kms)     &  ($'$)  & (dex)           & (\kms)  & (mas\,yr$^{-1}$) & (mas\,yr$^{-1}$) \\
\hline
0021      & B1 III        & \ltf  & 4.58  & 4.60  & 166   & 0.86$\pm$0.08  & -1.26$\pm$0.06  \\
0023      & B0.2: (Be-Fe) & 57    & 1.94  & 4.81  & 161   & 0.87$\pm$0.06  & -1.21$\pm$ 0.04  \\
0026      & B0 IV (Nstr)  & 68    & 3.87  & 4.79  & 222   & 0.59$\pm$0.07  & -1.18$\pm$0.05  \\
0043      & B0 V          & \ltf  & 4.02  & 4.71  & 172   & 0.73$\pm$0.10  & -1.17$\pm$0.07  \\
0044      & B1 II         & \ltf  & 3.25  & 4.33  & 146   & 1.01$\pm$0.08  & -1.22$\pm$0.06  \\
0047      & B2.5 III      & 55    & 9.35  & 4.15  & 137   & 0.91$\pm$0.09  & -1.23$\pm$0.06  \\
0054      & B1 V          & \ltf  & 2.88  & 4.47  & 164   & 0.84$\pm$0.07  & -1.38$\pm$0.05  \\
0056      & B0 V          & \ltf  & 1.06  & 4.55  & 178   & 0.93$\pm$0.09  & -1.18$\pm$0.07  \\
0057      & B2.5 III      & 80    & 5.38  & 4.08  & 143   & 0.62$\pm$0.11  & -1.21$\pm$0.07  \\
0062      & B0.2 V        & \ltf  & 6.61  & 4.45  & 136   & 0.99$\pm$0.10  & -1.19$\pm$0.07  \\
0089      & B1-2 (Be-Fe)  & 74    & 8.07  & 4.22  & 185   & 0.80$\pm$0.08  & -1.30$\pm$0.06  \\
0094      & B0.7 V        & \ltf  & 5.28  & 4.28  & 138   & 0.96$\pm$0.12  & -1.25$\pm$0.08  \\
0098      & B1.5 V        & 60    & 2.70  & 4.16  & 158   & 0.86$\pm$0.11  & -1.28$\pm$0.07  \\
0101      & B1 V          & \ltf  & 4.96  & 4.18  & 184   & 1.44$\pm$0.35  & -1.26$\pm$0.25  \\
0102      & B3 III        & \ltf  & 5.34  & 3.75  & 144   & -        & -                     \\
0103      & B0.5 V        & \ltf  & 7.54  & 4.26  & 137   & 0.84$\pm$0.12  & -1.19$\pm$0.08  \\
0111      & B0.5 V        & 49    & 0.49  & 4.23  & 163   & -     -  & -     -               \\
0116      & B1 V          & \ltf  & 6.26  & 4.15  & 154   & 0.74$\pm$0.12  & -1.33$\pm$0.09  \\
1028      & B2.5 III      & \ltf  & 7.75  & 4.38  & 169   & 1.06$\pm$0.11  & -1.07$\pm$0.08  \\
1029$^*$  & B2.5 III      & 55    & 6.48  & 4.38  & 123   & 0.83$\pm$0.09  & -1.07$\pm$0.06  \\
1036      & B2.5 III-II   & 57    & 4.39  & 4.31  & 146   & 0.85$\pm$0.08  & -1.27$\pm$0.06  \\
1041      & B2 III-II     & \ltf  & 8.29  & 4.37  & 181   & 0.72$\pm$0.10  & -1.21$\pm$0.07  \\
1050$^*$  & B1.5 II       & \ltf  & 7.02  & 4.44  & 185   & 0.74$\pm$0.06  & -1.33$\pm$0.04  \\
1053      & B0.5 III      & \ltf  & 6.44  & 4.50  & 178   & 0.30$\pm$0.10  & -1.13$\pm$0.07  \\
1068      & B0.7 III      & 60    & 5.97  & 4.35  & 185   & -        & -                     \\
1070      & B2 III        & \ltf  & 1.31  & 4.18  & 165   & 0.75$\pm$0.08  & -1.22$\pm$0.06  \\
1076$^*$  & B1 III        & 69    & 7.26  & 4.36  & 183   & 0.88$\pm$0.11  & -1.35$\pm$0.09  \\
1079      & B0 III        & \ltf  & 0.77  & 4.50  & 171   & 0.81$\pm$0.10  & -1.30$\pm$0.08  \\
1080      & B0.5 III      & \ltf  & 6.15  & 4.40  & 170   & 0.80$\pm$0.09  & -1.39$\pm$0.06  \\
1081$^*$  & B1 III        & \ltf  & 7.67  & 4.35  & 183   & 0.78$\pm$0.10  & -1.23$\pm$0.08  \\
1082      & B0 III        & \ltf  & 6.92  & 4.44  & 186   & 0.87$\pm$0.11  & -1.23$\pm$0.07  \\
1093      & B1 V          & \ltf  & 5.75  & 4.34  & 151   & 1.73$\pm$0.17  & -1.14$\pm$0.09  \\
1109      & B1.5 III      & \ltf  & 7.69  & 4.08  & 171   & 0.67$\pm$0.12  & -1.27$\pm$0.08  \\
1112      & B0 V          & 76    & 9.61  & 4.43  & 130   & 0.58$\pm$0.10  & -1.21$\pm$0.09  \\
1113      & B3 III        & 50    & 6.99  & 3.87  & 140   & 0.86$\pm$0.10  & -1.33$\pm$0.07  \\
1119$^*$  & B1 V          & \ltf  & 4.36  & 4.23  & 183   & 0.84$\pm$0.13  & -1.28$\pm$0.08  \\
1125$^*$  & B0 V          & \ltf  & 4.90  & 4.44  & 158   & 0.78$\pm$0.11  & -1.20$\pm$0.08  \\
1131      & B3 III        & \ltf  & 6.17  & 3.80  & 124   & 0.79$\pm$0.12  & -1.24$\pm$0.07  \\
1133      & B1 V          & \ltf  & 7.44  & 4.18  & 180   & 0.92$\pm$0.09  & -1.32$\pm$0.07  \\
1141$^*$  & B2 III        & \ltf  & 8.03  & 3.91  & 148   & 1.14$\pm$0.09  & -1.37$\pm$0.07  \\
1142$^*$  & B1.5 III      & \ltf  & 2.58  & 3.97  & 162   & 0.67$\pm$0.13  & -0.79$\pm$0.08  \\
1150$^*$  & B1 V          & 77    & 6.37  & 4.13  & 184   & 0.64$\pm$0.11  & -1.22$\pm$0.07  \\
1153$^*$  & B2 V          & 60    & 6.55  & 4.04  & 172   & 1.03$\pm$0.12  & -1.23$\pm$0.09  \\
1173$^*$  & B1 V          & 72    & 4.20  & 4.08  & 169   & 1.06$\pm$0.11  & -1.37$\pm$0.07  \\
1176      & B1.5 V        & \ltf  & 7.29  & 4.03  & 126   & 0.92$\pm$0.11  & -1.15$\pm$0.07  \\
1177      & B1-3e (Be-Fe) & 58    & 6.93  & 3.99  & 158   & 0.83$\pm$0.08  & -1.20$\pm$0.06  \\
1183$^*$  & B2.5 V        & \ltf  & 9.04  & 3.93  & 146   & 0.54$\pm$0.13  & -1.43$\pm$0.08  \\
1186      & B2 V          & 41    & 4.09  & 3.98  & 169   & 0.92$\pm$0.16  & -1.12$\pm$0.10  \\
1198      & B1.5 V        & 46    & 4.46  & 4.00  & 166   & 1.05$\pm$0.13  & -1.36$\pm$0.08  \\
1199      & B0 V          & \ltf  & 0.94  & 4.32  & 162   & 0.91$\pm$0.12  & -1.52$\pm$0.08  \\
1200      & B1 V          & \ltf  & 6.74  & 4.04  & 177   & 0.75$\pm$0.10  & -1.26$\pm$0.07  \\
1206      & B2.5 V        & 46    & 3.88  & 3.89  & 145   & 0.98$\pm$0.12  & -1.43$\pm$0.09  \\
1207      & B3 V          & \ltf  & 5.60  & 3.79  & 164   & 1.37$\pm$0.15  & -1.59$\pm$0.11  \\
1217      & B2.5 V        & \ltf  & 5.90  & 3.86  & 150   & 0.82$\pm$0.12  & -1.28$\pm$0.08  \\
1221      & B0.5 V        & \ltf  & 2.81  & 4.06  & 175   & 0.42$\pm$0.13  & -1.23$\pm$0.09  \\
1223      & B1 V          & 55    & 6.29  & 3.98  & 139   & 1.60$\pm$0.14  & -1.13$\pm$0.09  \\
\hline
\end{tabular}
\end{center}
\end{table*}
\setcounter{table}{0}

\begin{table*}
\caption{(continued.)}
\begin{center}
	\begin{tabular}{llrrccccc}
	\hline\hline
	Star   & Classification & \vsini     &  $r$    & $\log$\,L/\Lsun & \vr     & PM(RA) & PM(Dec)   \\ 
	&                       & (\kms)     &  ($'$)  & (dex)           & (\kms)  &  (mas\,yr$^{-1}$) & (mas\,yr$^{-1}$) \\
\hline
1225      & B3 V          & 79    & 5.84  & 3.75  & 167   & 0.69$\pm$0.13  & -1.17$\pm$0.09  \\
1227$^*$  & B3 V          & 68    & 8.54  & 3.75  & 180   & 1.01$\pm$0.13  & -1.25$\pm$0.10  \\
1233      & B0.2 V        & \ltf  & 3.22  & 4.10  & 159   & 1.08$\pm$0.12  & -1.24$\pm$0.08  \\
1234      & B2.5 V        & \ltf  & 2.92  & 3.84  & 132   & 0.49$\pm$0.16  & -0.54$\pm$0.11  \\
1236$^*$  & B1 V          & 67    & 1.17  & 3.96  & 174   & 0.91$\pm$0.14  & -1.25$\pm$0.10  \\
1239      & B2 V          & 41    & 7.37  & 3.88  & 151   & 0.87$\pm$0.13  & -1.28$\pm$0.10  \\
1243$^*$  & B3 V          & 70    & 4.02  & 3.73  & 160   & 0.77$\pm$0.13  & -1.26$\pm$0.08  \\
\hline
\end{tabular}
\tablefoot{Radial distances ($r$) in col. 5 are in arcmin from \#1001
  ($\alpha$\,$=$\,00\,59\,04.49, $\delta$\,$=$\,$-$72\,10\,24.7,
  J2000), where 1$'$\,$\approx$\,17.5\,pc.}
\end{center}
\end{table*}

\section{Atmospheric parameters} \label{s_atm}

All the results presented here, including those from previous investigations, have employed model atmosphere grids calculated with the {\sc tlusty} and {\sc synspec} codes \citep{hub88, hub95, hub98, lan07}.  They cover a range of effective temperature, 12\,000K $\leq$\teff$\leq$35\,000K in steps of typically 1\,500K. Logarithmic gravities (in cm s$^{-2}$) range from 4.5 dex down to the Eddington limit in steps of 0.25 dex, and microturbulences are from 0-30 \kms\ in steps of 5 \kms. As discussed in  \citet{rya03} and \citet{duf05}, equivalent widths and line profiles interpolated within these grids are in good agreement with those calculated explicitly at the relevant atmospheric parameters.

These non-LTE codes adopt the `classical' stationary model atmosphere assumptions, that is plane-parallel geometry, hydrostatic equilibrium, and that the optical spectrum is unaffected by winds. As the targets considered here have luminosity classes III to V, such an approach should provide reliable results. The grids assumed a normal helium to hydrogen ratio (0.1 by number of atoms) and the validity of this is discussed later. Grids have been calculated for a range of metallicities with that for an SMC metallicity being used here. 

Abundance estimates are provided for both nitrogen and magnesium in Table \ref{t_Atm1} and Table \ref{t_Atm2}, whilst possible variations in the carbon abundance estimates are considered in Sect.\ \ref{d_N}. The first utilised a 51 level \ion{N}{ii} ion with 280 radiative transitions \citep{all03} and a 36 level \ion{N}{iii} ion with 184 radiative transitions \citep{lan03}, together with the ground states of the \ion{N}{i} and \ion{N}{iv} ions. As discussed in Paper I, predicted non-LTE effects for \ion{N}{ii} are relatively small. For example, the difference between the nitrogen abundance estimates deduced using LTE and non-LTE methodologies for \teff=25\,000K and \logg=4.0 dex are of the order of 0.15\,dex for the line at 3995\AA. Our \ion{Mg}{ii} ion had 31 levels with 184 radiative transitions \citep{all03, lan07}, together with the ground state of the \ion{Mg}{iii} ion. As first discussed by \citet{mih72}, non-LTE effects in \ion{Mg}{ii} are small for B-type main sequence stars. For example, for \teff=25\,000K, \logg=4.0 dex and an SMC metallicity, non-LTE effects change the abundance estimated from the \ion{Mg}{ii} doublet at 4481\,\AA\ by only 0.03\,dex.
	
For carbon, a 40 level \ion{C}{ii} ion with 189 radiative transitions \citep{all03} and a 23 level \ion{C}{iii} ion with 258 radiative transitions \citep{lan03} was adopted together with the ground states of the \ion{C}{i} and \ion{C}{iv} ions.  As discussed by \citet{sig96}, the \ion{C}{ii} spectrum has been found to be subject to significant non-LTE effects, requiring sophisticated model ions \citep[see, for example,][]{nie06, ale19}.  Indeed \citet{hun07}, using the model atmosphere grids adopted here, found that the carbon abundances estimates for targets in both the Galaxy and the Magellanic Clouds were lower than those found from \ion{H}{ii} regions; however the {\em relative} abundance estimates deduced from these grids appeared to be reliable.Further information on the {\sc tlusty} grids are provided in \citet{rya03} and \citet{duf05}.

For the FSMS targets, atmospheric parameters -- effective temperature, \teff; logarithmic gravity, \logg; and microtubulent velocity, \vt\ -- have been taken directly from the previous analyses of \citet{hun07, hun08a} and \citet{dun11}, apart from those for \#0111 for which no estimates were available. The reason for the omission of this star from the previous analyses is unclear as it has been classified as B0.5\,V in \citet{eva06}. Its FSMS spectroscopy was analysed in Paper I and the results have been incorporated into Table \ref{t_Atm2}. Two targets, \#0023 and \#0089 had been classified as Be-type by \citet{eva06} and subsequently analysed by \citet{dun11}. These targets have been included in our analysis but the results should be treated with caution given the complex nature of their spectra \citep[see, for example][]{ahm17}.

For the MSC targets, effective temperatures and logarithmic gravities were taken directly from Paper II. We have estimated microturbulences from the \ion{Si}{iii} multiplet near 4560\AA\ using two different methods viz. removing the variation of the abundance estimate with line strength (method 1) and reproducing the SMC silicon abundance (method 2). For the latter we adopted a baseline silicon abundance of 6.8 dex consistent with the estimates for B-type stars of \citet[][6.80\,dex]{hun07}, young late-type stars of \citet[][6.87\,dex]{gon99}, K-type stars of \citet[][6.81\,dex]{hil97b} and F-type stars from \citet[][6.74\,dex]{luc98}; the latter two investigations gave estimates relative to the solar abundance and this has been assumed to be 7.51\,dex \citep{asp09}.

Nine targets had a maximum silicon abundance estimate (i.e.\ for \vt\ = 0 \kms) that was more than 0.1\,dex below the adopted SMC value and these are identified in Table \ref{t_Atm1} and \ref{t_Atm2}. Additionally it was not always possible to remove the variation of silicon abundance estimate with line strength. Similar effects have been found by \citet{duf18}, \citet{mce14} and \citet{hun07}, with the last discussing the possible explanations in some detail. One plausible explanation is that these are unidentified binaries and we discuss this further in Sect. \ref{d_N}. For these targets, we have adopted the best estimate, that is zero microturbulence. The two methods yielded similar estimates for the microturbulences with the mean difference (method 1 -- method 2) being $-$0.1$\pm$1.9\,\kms. We have adopted the estimates based on the silicon abundance in order to be consistent with the FSMS estimates but note that our principle conclusions would be unaffected if we have chosen the estimates from the other methodology.

Spectroscopy of two MSC stars has not been analysed. Firstly, given the problems highlighted earlier, we were unable to obtain a reliable upper limit for the nitrogen abundance in the Be-type spectrum of \#1177. Secondly, \#1225 had independent \vsini\ estimates of 79 and 96\,\kms\ (see Paper I for details of the different methodologies) and hence lay near to the upper \vsini\ limit for our sample. Further inspection of the metal lines in its spectra indicated that its projected rotational velocity was probably greater than 80\,\kms\ and it was therefore excluded from the analysis. 

The adopted atmospheric parameters are summarised in Tables \ref{t_Atm1} and \ref{t_Atm2}. The former is for the targets with \vsini$\leq$40\,\kms\ and the latter for the remaining targets. The stochastic uncertainties in the effective temperature and logarithmic gravity have been estimated by \citet{hun07,hun08a,346_lr} to be of the order of 1\,000\,K and 0.2\,dex respectively. However for MSC targets where the effective temperatures were estimated from the spectral type the former could be larger and was estimated in Paper II to be 1\,500\,K. For the microturbulent velocity estimates, we have adopted a stochastic uncertainty of $\pm$3\,\kms, consistent with the differences in the estimates from our two methods.

\begin{table*}
\begin{center}
\caption{Atmospheric parameters and element abundances for the targets with a projected rotational velocity, \vsini$\leq$40\,\kms. The methods for estimating the atmospheric parameters were as follows -- H08: taken from \citet{hun08a}; A: adopted from effective temperature scales of \citet{tru07}; E: extrapolated from these temperature scales; Si: silicon ionization balance;  He+: \ion{He}{ii} lines at 4541 and 4686\AA. Further details can be found in Paper I. Targets where the silicon abundance estimate was less than 6.8\,dex are marked with an asterisk. Also listed are the spectral types taken from Table \ref{t_Targets} and the stellar mass and age estimates taken from Paper I. For the latter, this has also been estimated as a fraction of the main sequence lifetime.}\label{t_Atm1}
\begin{tabular}{llclccccrrr}
\hline\hline
Star   & Classification   & \teff  & Method  & \logg   & \vt    &  \MgtoH & \NtoH & Mass   
& Age  & Age\\
		&          &   (K)  &         & (\cms)  & (\kms) &  (dex)  & (dex) & (\Msun)
& (Myr) & (TAMS)\\
\hline
0021    &  B1 III    &  25150  &  H08  &  3.50  &  3  &  6.76  &  6.88  &  15.0  &  10.0  &  0.82  \\
0043$^*$&  B0 V      &  33000  &  H08  &  4.25  &  0  &  6.83  &  6.84  &  17.0  &  6.2   &  0.56  \\
0044$^*$&  B1 III    &  23000  &  H08  &  3.50  &  4  &  6.66  &  6.68  &  10.6  &  17.2  &  0.93  \\
0054    &  B1 V      &  29000  &  H08  &  4.30  &  0  &  6.45  &  <6.8  &  13.2  &  9.8   &  0.71  \\
0056    &  B0 V      &  31000  &  H08  &  3.80  &  1  &  6.81  &  7.51  &  14.6  &  7.8   &  0.63  \\
0062    &  B0.2 V    &  29750  &  H08  &  4.00  &  6  &  6.75  &  7.27  &  13.4  &  9.0   &  0.66  \\
0094    &  B0.7 V    &  28500  &  H08  &  4.00  &  6  &  6.75  &  7.36  &  11.8  &  10.5  &  0.66  \\
0101    &  B1 V      &  27300  &  H08  &  4.25  &  0  &  6.49  &  6.73  &  11.0  &  12.4  &  0.71  \\
0102$^*$&  B3 III    &  17700  &  H08  &  3.70  &  3  &  6.45  &  <7.3  &  6.8   &  40.9  &  0.98  \\
0103    &  B0.5 V    &  29500  &  H08  &  4.00  &  1  &  6.82  &  7.54  &  12.2  &  8.8   &  0.58  \\
0116    &  B1 V      &  28250  &  H08  &  4.10  &  0  &  6.70  &  6.71  &  11.2  &  10.4  &  0.61  \\
1028    &  B2.5 III  &  22000  &  Si   &  3.45  &  1  &  6.72  &  7.12  &  10.0  &  19.9  &  0.98  \\
1041    &  B2 III-II &  22500  &  Si   &  3.35  &  4  &  6.61  &  6.52  &  10.4  &  18.8  &  0.99  \\
1050    &  B1.5 II   &  24100  &  Si   &  3.45  &  3  &  6.77  &  7.07  &  11.6  &  14.6  &  0.90  \\
1053    &  B0.5 III  &  27500  &  Si   &  3.60  &  4  &  6.91  &  7.59  &  12.8  &  11.8  &  0.83  \\
1070    &  B2 III    &  21200  &  A    &  3.40  &  3  &  6.40  &  6.71  &  9.4   &  22.1  &  0.98  \\
1079$^*$&  B0 III    &  31000  &  He+  &  4.00  &  0  &  --    &  <7.2  &  14.2  &  7.7   &  0.60  \\
1080    &  B0.5 III  &  27800  &  Si   &  3.90  &  3  &  6.83  &  <6.9  &  12.4  &  11.3  &  0.76  \\
1081    &  B1 III    &  26400  &  Si   &  3.75  &  1  &  6.62  &  7.08  &  11.5  &  13.1  &  0.80  \\
1082$^*$&  B0 III    &  29000  &  Si   &  3.85  &  0  &  6.93  &  <7.1  &  13.0  &  9.8   &  0.70  \\
1093    &  B1 V      &  26000  &  He+  &  4.00  &  0  &  6.83  &  6.87  &  11.8  &  12.0  &  0.76  \\
1109    &  B1.5 III  &  24000  &  Si   &  3.60  &  4  &  7.04  &  7.05  &  9.0 	 &  21.5  &  0.88  \\
1119    &  B1 V      &  23000  &  Si   &  3.60  &  5  &  6.56  &  6.61  &  11.2  &  12.4  &  0.73  \\
1125    &  B0 V      &  33000  &  He+  &  4.40  &  4  &  --    &  7.73  &  15.0  &  4.6   &  0.38  \\
1131    &  B3 III    &  18450  &  A    &  3.55  &  0  &  6.58  &  <7.1  &  7.0   &  36.7  &  0.93  \\
1133    &  B1 V      &  27300  &  A    &  4.00  &  3  &  7.12  &  <6.8  &  11.0  &  12.4  &  0.71  \\
1141    &  B2 III    &  21200  &  A    &  3.75  &  6  &  6.72  &  6.44  &  8.0   &  26.7  &  0.88  \\
1142    &  B1.5 III  &  22550  &  A    &  3.70  &  4  &  6.65  &  6.91  &  8.6   &  22.5  &  0.85  \\
1176$^*$&  B1.5 V    &  26100  &  A    &  4.35  &  0  &  6.66  &  6.63  &  9.8   &  14.6  &  0.69  \\
1183$^*$&  B2.5 V    &  23900  &  E    &  4.35  &  0  &  6.68  &  <6.8  &  8.8   &  19.9  &  0.78  \\
1199    &  B0 V      &  34750  &  He+  &  4.40  &  5  &  6.56  &  <8.0  &  15.6  &  0.0   &  0.00  \\
1200    &  B1 V      &  27300  &  A    &  4.10  &  3  &  6.86  &  6.66  &  10.2  &  11.9  &  0.61  \\
1207    &  B3 V      &  21500  &  E    &  4.25  &  5  &  6.40  &  <7.0  &  7.6   &  27.2  &  0.81  \\
1217$^*$&  B2.5 V    &  23900  &  E    &  4.15  &  0  &  6.66  &  <6.8  &  8.4   &  20.0  &  0.72  \\
1221    &  B0.5 V    &  29200  &  Si   &  4.25  &  3  &  7.07  &  <7.0  &  11.2  &  6.8   &  0.40  \\
1233    &  B0.2 V    &  30650  &  Si   &  4.40  &  0  &  6.89  &  <7.2  &  12.0  &  3.0   &  0.19  \\
1234$^*$&  B2.5 V    &  23900  &  E    &  4.00  &  0  &  6.82  &  <6.8  &  8.4   &  20.0  &  0.72  \\
\hline
\end{tabular}
\end{center}
\end{table*}

\begin{table*}
\begin{center}
  \caption{Atmospheric parameters and element abundances for the
    targets with a projected rotational velocity,
    40$<$\vsini$\leq$80\,\kms. See the Table \ref{t_Atm1} caption for more
    details, with the addition that in col. 4, D11 indicates estimates
    from \citet{dun11}.}\label{t_Atm2}
\begin{tabular}{llclccccrrr}
\hline\hline
Star   & Classification  & \teff  & Method  & \logg   &  \vt    &  \MgtoH & \NtoH & Mass    & Age & Age   \\
&                 &   (K)  &         & (\cms)  & (\kms)  &  (dex)  & (dex) & (\Msun) & (Myr) & (TAMS) \\
\hline
0023    &  B0.2: (Be-Fe) &   31000   &  D11 &   3.65    &   10  &    -      &   7.52  &  17.2 &  7.3   &   0.66    \\
0026    &  B0 IV (Nstr)  &   31000   &  H08 &   3.70    &   5   &   6.65    &   7.76  &  17.0 &  7.3   &   0.66    \\
0047    &  B2.5 III      &   19850   &  H08 &   3.25    &   2   &   6.74    &   7.34  &  9.0  &  23.7  &   0.97    \\
0057    &  B2.5 III      &   19850   &  H08 &   3.35    &   4   &   6.53    &   <6.8  &  8.6  &  25.6  &   0.96    \\
0089    &  B1-2 (Be-Fe)  &   25500   &  D11 &   4.10    &   0   &    --     &   <7.1  &  10.8 &  13.9  &   0.77    \\
0098    &  B1.5 V        &   26100   &  H08 &   4.05    &   1   &   6.99    &   6.87  &  10.4 &  14.2  &   0.75    \\
0111    &  B0.5 V        &   28000   &  He+ &   4.10    &   0   &   --      &   <7.0  &  12.2 &  8.1   &   0.54    \\
1029    &  B2.5 III      &   19850   &  A   &   3.25    &   1   &   6.59    &   6.57  &  10.0 &  19.9  &   0.98    \\
1036    &  B2.5 III-II   &   19850   &  A   &   3.25    &   3   &   6.61    &   7.85  &  9.80 &  20.70 &   0.98    \\
1068    &  B0.7 III      &   25500   &  He+ &   3.85    &   2   &   6.87    &   <6.8  &  11.4 &  14.2  &   0.86    \\
1076    &  B1 III        &   26500   &  Si  &   3.90    &   1   &   6.73    &   <6.6  &  11.8 &  12.7  &   0.80    \\
1112    &  B0 V          &   32100   &  Si  &   4.20    &   2   &   - -     &   <7.2  &  14.7 &  5.8   &   0.47    \\
1113    &  B3 III        &   18450   &  A   &   3.30    &   4   &   6.55    &   7.31  &  7.4  &  33.1  &   0.93    \\
1150    &  B1 V          &   26500   &  He+ &   4.10    &   2   &   6.77    &   6.53  &  10.8 &  11.9  &   0.66    \\
1153    &  B2 V          &   24950   &  A   &   4.30    &   0   &   6.89    &   <6.8  &  9.6  &  16.7  &   0.77    \\
1173    &  B1 V          &   26000   &  A   &   4.10    &   3   &   6.84    &   7.37  &  10.0 &  11.7  &   0.58    \\
1186$^*$&  B2 V          &   24950   &  A   &   4.05    &   0   &   7.03    &   <6.8  &  9.4  &  16.3  &   0.72    \\
1198    &  B1.5 V        &   26000   &  A   &   3.80    &   0   &   6.98    &   7.46  &  9.8  &  14.2  &   0.68    \\
1206    &  B2.5 V        &   23900   &  E   &   4.20    &   1   &   6.93    &   <6.8  &  8.6  &  19.3  &   0.73    \\
1223    &  B1 V          &   27300   &  A   &   4.30    &   0   &   6.87    &   <6.9  &  10.2 &  10.8  &   0.55    \\
1227$^*$&  B3 V          &   21500   &  E   &   4.05    &   0   &   6.53    &   <7.1  &  7.4  &  27.5  &   0.78    \\
1236    &  B1 V          &   27000   &  A   &   4.30    &   3   &   6.68    &   <6.9  &  10.2 &  9.0   &   0.46    \\
1239    &  B2 V          &   24950   &  A   &   4.35    &   0   &   6.92    &   <7.0  &  8.8  &  16.8  &   0.66    \\
1243    &  B3 V          &   21500   &  E   &   4.15    &   0   &   6.91    &   <7.3  &  7.4  &  27.8  &   0.78    \\
\hline
\end{tabular}
\end{center}
\end{table*}

As a test of the validity of our adopted atmospheric parameters, we have also listed magnesium abundance estimates (where \MgtoH=$\log$ [Mg/H]+12) for all our sample in Tables \ref{t_Atm1} and \ref{t_Atm2}. For the FSMS targets, these were taken directly from \citet{hun07, hun08a}\footnote{No estimates were available for the two FSMS targets analysed by \citet{dun11}}, whilst for the MSC targets, they were estimated from the \ion{Mg}{ii} doublet at 4481\,\AA. They have a mean of 6.74$\pm$0.17\, dex, in excellent good agreement with SMC estimates found from B-type stars \citep[][6.75 and 6.78\,dex]{hun07,kor00}, A-type stars \citep[][6.82\,dex]{ven99} and young late-type stars \citep[][6.73\,dex]{gon99}.

\section{Nitrogen Abundances} \label{s_N}

Nitrogen abundances (where \NtoH=$\log$ [N/H]+12) were estimated primarily using the singlet transition at 3995\,\AA\ as this feature was the strongest \ion{N}{ii} line in the wavelength regions covered by both observational configurations and appeared unblended.  For stars where this line was not visible in the spectroscopy, an upper limit on its equivalent width was estimated from the S/N as discussed in Sect. \ref{s_obs}. The corresponding abundance estimates (and upper limits) are summarised in Table \ref{t_Atm1} and \ref{t_Atm2}.

The \ion{N}{ii} spectrum in the LR02 and LR03 wavelength regions has been discussed in detail in Paper I. Other lines are intrinsically weaker and often prone to blending. However we have searched the spectra of all of our targets for other \ion{N}{ii} features that were included in our {\sc tlusty} model atmosphere grids. Convincing identifications were obtained for nine targets, generally those that appear to have enhanced nitrogen abundances. The corresponding abundance estimates are summarised in Table \ref{t_N}, with values for the line at 3995\,\AA\ being taken directly from Tables \ref{t_Atm1} and \ref{t_Atm2}.

In general there is excellent agreement between the estimates from the \ion{N}{ii} line at 3995\AA\ and those from the other lines. Indeed the mean estimates always agree to within 0.05 dex or better with the former. This is consistent with the nitrogen abundance estimates for the VFTS targets in Paper I, where good agreement was also found. For consistency, we have adopted the estimates from the line at 3995\AA\  but note that our conclusions would remain unchanged if we had adopted the estimates from Table \ref{t_N} where available.

\begin{table*}
\begin{center}
\caption{Nitrogen abundance estimates from different lines. The values for the line at 3995\,\AA\ have been taken directly from Tables \ref{t_Atm1} and \ref{t_Atm2}. For the FSMS targets (those with identifications of less than \#1000) the spectroscopy did not include the \ion{N}{ii} line at 5001\AA.}\label{t_N}
\begin{tabular}{lcccccccc}
\hline\hline
Star &  \multicolumn{8}{c}{Nitrogen Abundances (dex)} \\ 
     & $\lambda$3995 & $\lambda$4447 & $\lambda$4601 & $\lambda$4613  & 
$\lambda$4621 & $\lambda$4630  & $\lambda$5001 & Mean  \\
\hline
0021 & 6.88 & 6.83 &    - & 6.92  &    - & 6.92  &    - & 6.89$\pm$0.04 \\
0026 & 7.76 &    - &    - &    -  &    - & 7.83  &    - & 7.79$\pm$0.05 \\
0047 & 7.34 & 7.35 & 7.39 & 7.36  & 7.18 & 7.42  &    - & 7.34$\pm$0.08 \\
0056 & 7.51 & 7.36 &    - &    -  &    - & 7.50  &    - & 7.46$\pm$0.08 \\
0094 & 7.36 &    - &    - &    -  &    - & 7.35  &    - & 7.35$\pm$0.01 \\
0103 & 7.54 & 7.64 &    - &    -  &    - & 7.50  &    - & 7.56$\pm$0.07 \\
1036 & 7.85 & 7.71 & 7.71 & 7.82  & 7.97 & 7.95  & 7.97 & 7.85$\pm$0.11 \\
1053 & 7.59 & 7.51 & 7.76 & 7.75  & 7.62 & 7.64  & 7.32 & 7.60$\pm$0.15 \\
1198 & 7.46 &    - &    - &    -  &    - & 7.56  &    - & 7.51$\pm$0.07 \\
   \hline  
\hline
\end{tabular}
\end{center}
\end{table*}

All these abundance estimates will be affected by uncertainties in the
atmospheric parameters which has been discussed by, for example,
\citet{hun07} and \citet{fra10}. Using a similar methodology, and
taking into account uncertainties in both the atmospheric parameters
and the observational data, we estimate a typical uncertainty of
0.2-0.3~dex on the estimated nitrogen abundance. However, due to the
use of a similar methodology for all targets, we would expect the
uncertainty in {\em relative} nitrogen abundances to be smaller. This
would be consistent with the relatively small standard deviation for
the magnesium abundance estimates found in Sect. \ref{s_atm}.

\section{Masses and Lifetimes}\label{s_M_A}

In Paper II, current stellar masses and ages were estimated using  {\sc bonnsai}\footnote{The {\sc bonnsai} web-service is available at: \newline \url{www.astro.uni-bonn.de/stars/bonnsai}.}, which uses a Bayesian methodology and the grids of models from \citet{bro11a} to constrain the evolutionary status of a given star \citep[see][for details]{sch14}. The SMC metallicity grid of models, a \citet{sal55} initial mass function, the initial rotational velocity distribution estimated by \citet{hun08a}, a random orientiation of spin axes, and a uniform age distribution were adopted as independent prior functions. The estimates of effective temperature, luminosity and projected rotational velocity were then used to deduce current masses and ages, which are listed in Tables~\ref{t_Atm1} and \ref{t_Atm2}. The former were very similar to the initial mass estimates with differences $<$5\%. 

The stellar mass estimates cover a range from approximately 7--17\Msun\ and the targets will therefore have a range of main sequence lifetimes. We have used the SMC models of \citet{bro11a} to estimate, the time required to reach the terminal age main sequence (TAMS) -- here defined as at a gravity, \logg=3.2\,dex (see Sect.\ \ref{s_obs}). These TAMS ages depend on the adopted initial rotational velocity and, for example, increase by typically 5\% as the rotational velocity increase from 0 to 400\,\kms. Given the low \vsini\ estimates for our targets (see also the discussion in Sect.\ \ref{d_freq}), we have adopted ages for the non-rotating models. In turn these have been used to deduce the stellar ages as a fraction of the TAMS age and these are also listed Tables~\ref{t_Atm1} and \ref{t_Atm2}.

{\sc bonnsai} also returned 1$\sigma$-uncertainties for all the quantities that it had estimated based on  the adopted errors in the effective temperature, luminosity and projected rotational velocities. For the stellar masses, the error estimates were relatively symmetric and adopting the larger values led to error estimates of 6-8\% for all our targets. For the age estimates, the errors were more asymmetric and again adopting the larger value led to estimates of 10-22\% for more than 80\% of our sample. The remaining targets had small age estimates (leading to the large percentage error estimates) with a median error estimate of 2.8\,Myr and a maximum of 4.5 Myr.  There may be additional uncertainties due to undetected binarity or line-of-sight composites (unresolved at the distance of the SMC). Simulations presented in Paper I implied that masses could be overestimated by up to 10\%, whilst age estimates both decreased and increased by up to 15\%.

\section{Discussion}
\subsection{Nitrogen abundances} \label{d_N}

SMC nitrogen abundances have been estimated from \ion{H}{ii} regions \citep{duf77, duf82, kur98, gar99, rey14}, early-type stars \citep{rol03, tru07, hun08a}, A-type stars \citep{ven99}, late-type stars \citep{rus89, luc92, hil97a} and planetary nebulae \citep{dop91, lei06}. For the stellar and PN estimates, the progenitor nitrogen abundance may have been modified by nucleosynthetic processes during their evolution, whilst the \ion{H}{ii} region estimates should be less affected. \citet{hun07} found a range of nitrogen abundances from their spectroscopic analysis of 14 SMC B-type stars and from their lowest estimates inferred a baseline abundance of \NtoH$\simeq$6.5 dex. This was subsequently adopted in the grid of SMC evolutionary models calculated by \citet{bro11a}. The \ion{H}{ii} region analyses listed above lead to similar estimates with, for example, the investigation by \citet{rey14} yielding a baseline nitrogen abundance estimate of 6.55 dex; additionally their review of the literature led to an estimate of 6.49 dex. Here, we will adopt a baseline abundance of 6.5\,dex  but discuss in Sect. \ref{d_sim} the implications of this choice.
 
The estimated nitrogen abundances summarised in Tables~\ref{t_Atm1} and \ref{t_Atm2} range from baseline to an enhancement of greater than 1.0 dex. Approximately 36\% have enhancements of less than 0.3 dex, with an additional 28\% of the sample have upper limits greater than 6.8 dex. Hence, approximately 64\% of our sample {\em could} have nitrogen enhancements of less than a factor of two. In Paper I, approximately three quarters of a similar B-type sample in the LMC were consistent with similarly small nitrogen abundance enhancements. These would arise if the targets had small rotational velocities and had evolved without any significant rotational mixing \citep[see, for example,][]{mae87, heg00b, mae09,  fri10}. 

The remaining SMC targets show significant enhancements in nitrogen with approximately 36\%, 23\% and 13\% having an enhancement of greater than 0.3, 0.6 and 0.9 dex respectively (see Table \ref{t_N_sim}). These could be rapidly-rotating single stars that have undergone rotational mixing and are being observed at small angles of inclination. However as discussed in Paper I, such cases would be rare. For example, an equatorial velocity of more than 270 \kms\ would be required to achieve a nitrogen enhancement of 0.9 dex (see Table \ref{t_v_e}). Assuming random axes of inclination, such a star would be observed to have a value of \vsini\,$\leq$40 \kms\ only about 1\% of the time. Additionally, the probability that such a star would be observed with 40\,$<$\,\vsini\,$\leq$\,80 \kms\ is approximately four times that of it being observed with \vsini\,$\leq$\,40 \kms. As discussed in Paper I, similar ratios are found for other values of the rotational velocity and just reflects the greater solid angle of inclination available to stars in the larger projected rotational velocity bin. By contrast, the nitrogen-rich targets in our samples appear just as likely to have \vsini\,$\leq$\,40 \kms\ as 40\,$<$\,\vsini\,$\leq$\,80 \kms, as can be seen from Table \ref{t_N_sim}. This is again consistent with the LMC results discussed in Paper I and provides strong observational evidence that not all these objects can be fast rotating single stars that have undergone rotational mixing. In Sect. \ref{d_sim}, we investigate this further by undertaking simulations to estimate the number of fast rotating stars that we might expect in our sample. 

\begin{figure}
	\epsfig{file=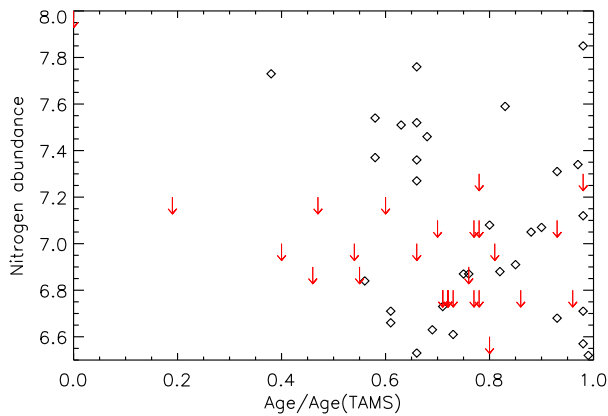,width=9cm, angle=0}\\
	\caption{Stellar nitrogen abundance estimates (and upper limits) plotted against the ratios of estimated ages to  TAMS ages.}
	\label{f_n_t}
\end{figure}

In Fig. \ref{f_n_t}, we plot the nitrogen abundance estimates (and upper limits) against the estimated ages as a fraction of the TAMS age. There is no evidence of any correlation between the two as might be expected if the nitrogen enhancements were due to rotational mixing \citep[see, for example][]{lyu04}. However the majority of our targets lie towards the end of their main sequence phase, whilst SMC evolutionary models \citep{bro11a, geo13} imply that significant nitrogen enrichment can occur at a relatively early stage of this phase.

For our nitrogen enriched stars, changes in the surface carbon and oxygen abundances are predicted by evolutionary models of single stars that include rotation. Indeed, \citet{mae14} argued that comparison of the relative N/C and N/O abundances would provide a powerful diagnostic of rotational mixing. The SMC models of \citet{bro11a} indicate that for a nitrogen enhancement of $\sim$0.9 dex (exhibited by 8 stars in our sample), the carbon and oxygen abundances would be changed by $-$0.1--0.5 dex and -0.03--0.08 dex respectively for a range of initial masses from 7--15\Msun. The smaller carbon depletions (and corresponding larger oxygen depletions) occur at lower mass and larger initial rotational velocity.
 
\begin{figure}
 \epsfig{file=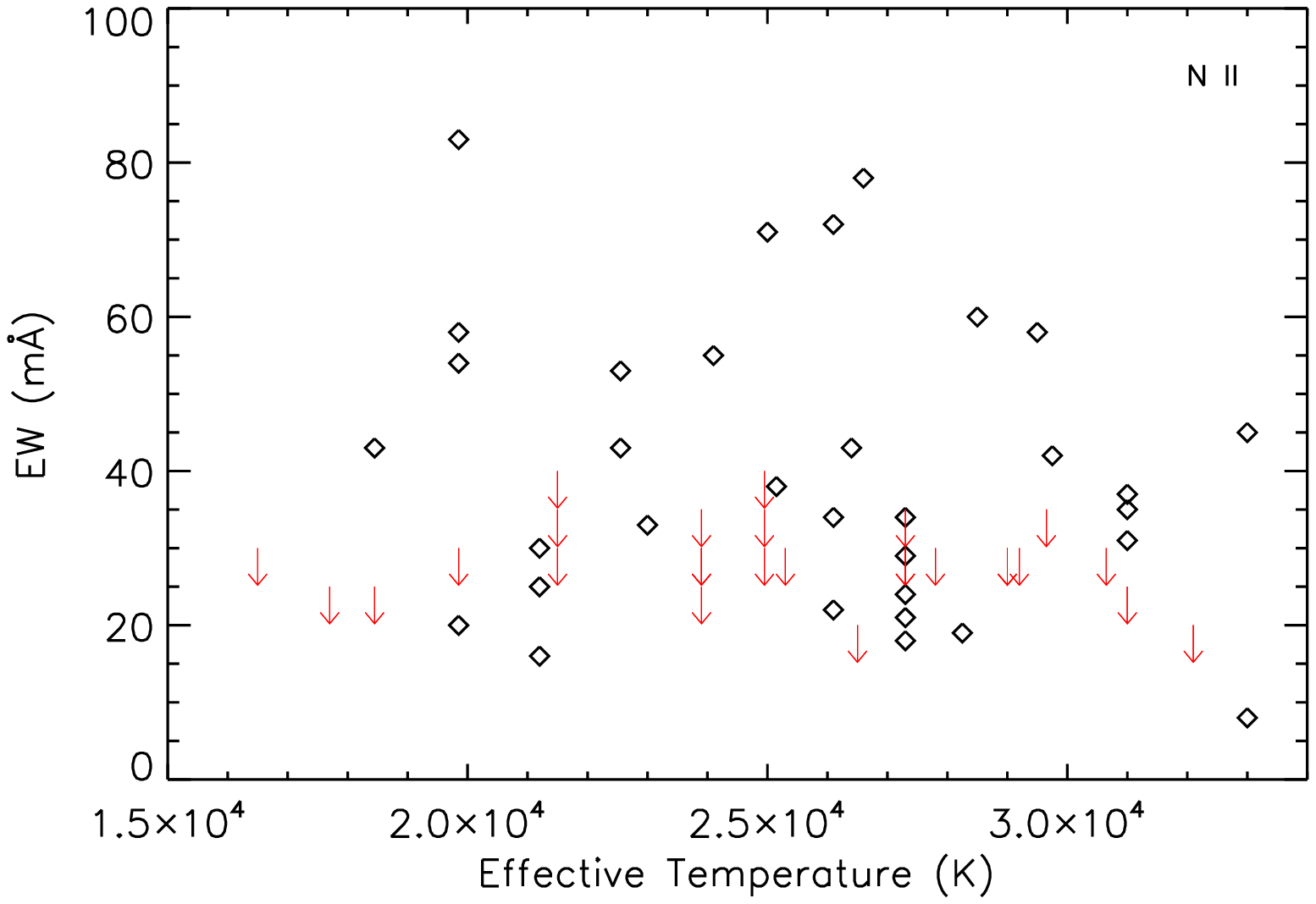,width=9cm, angle=0}\\
 \epsfig{file=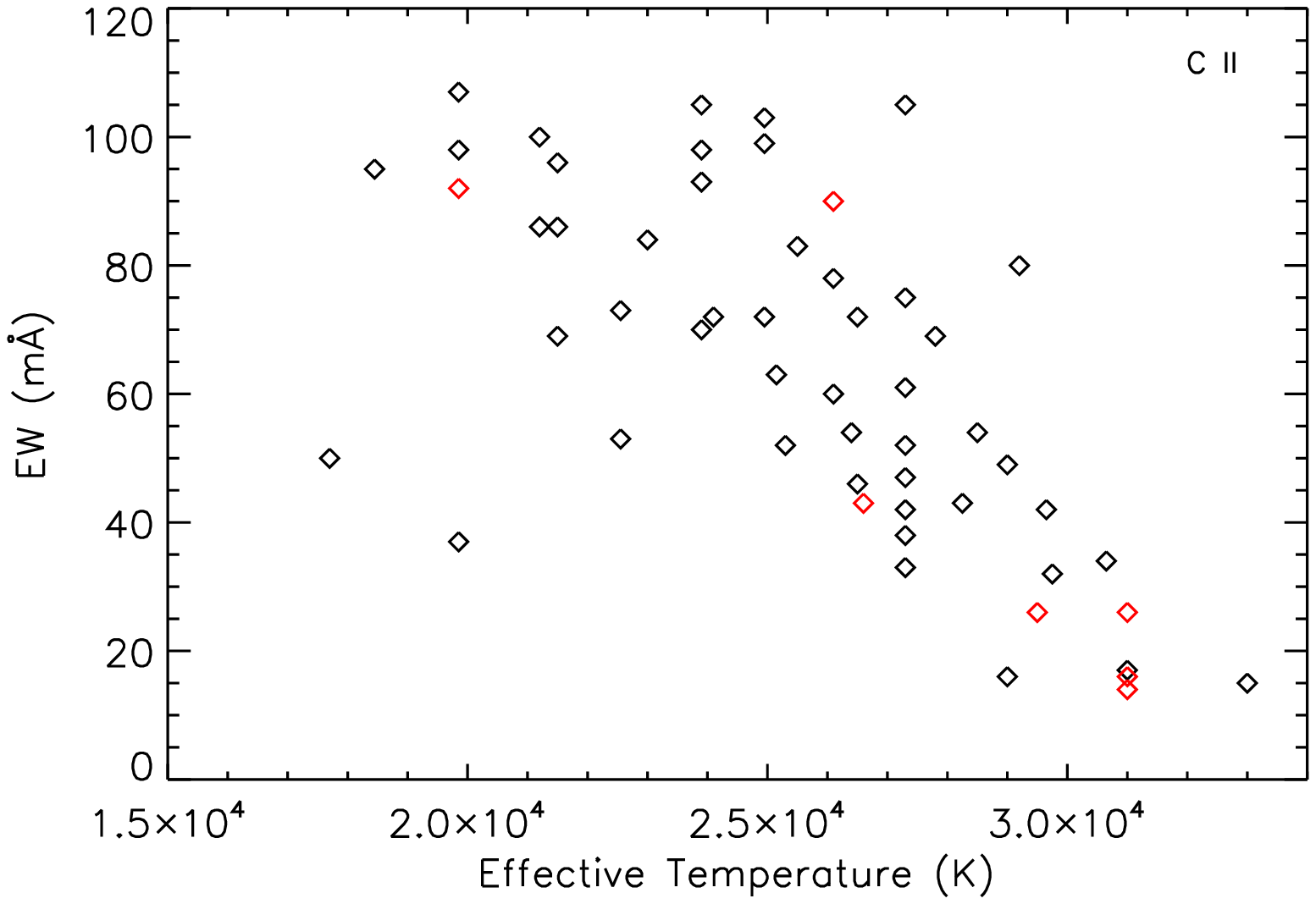,width=9cm, angle=0}\\
 \epsfig{file=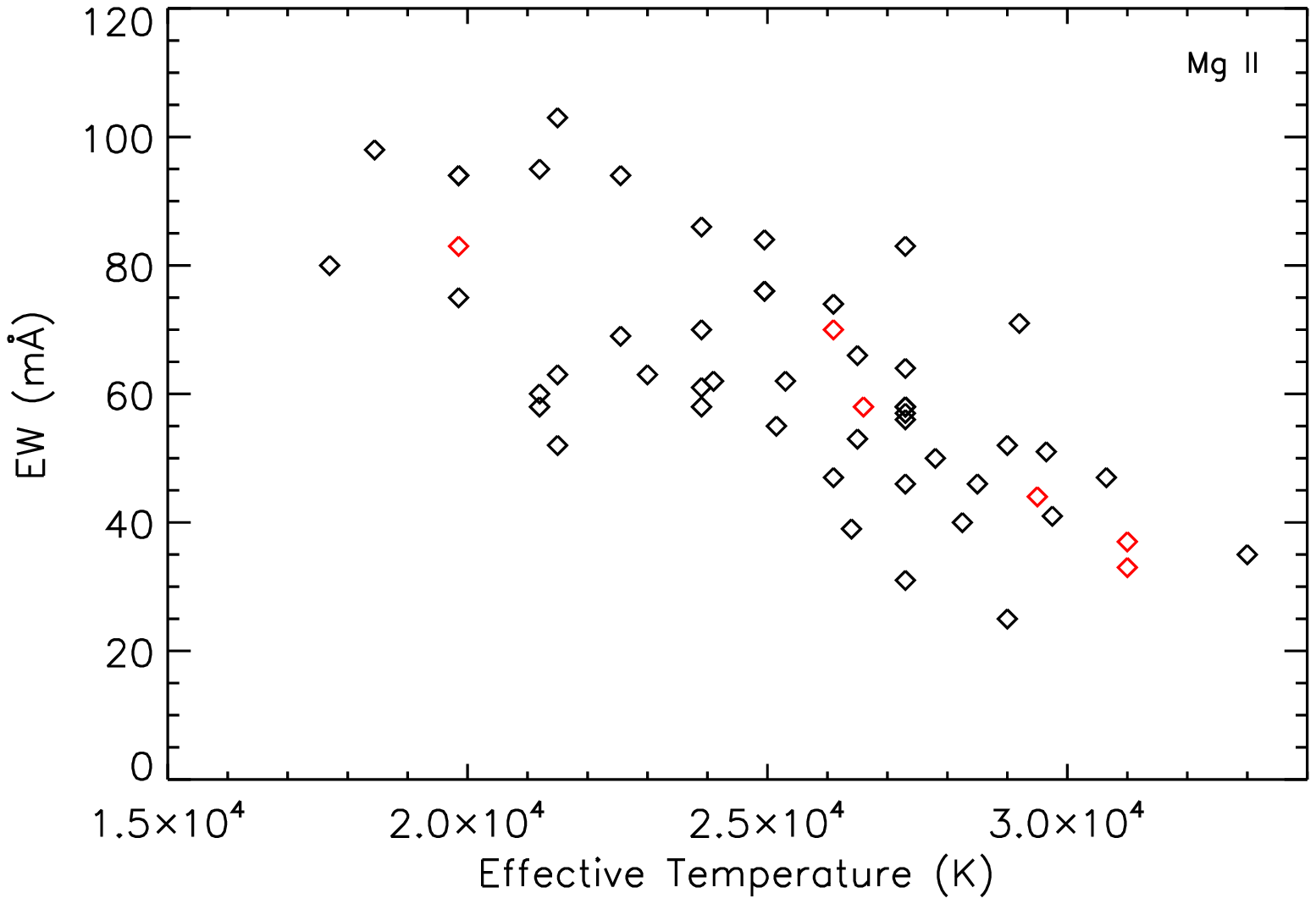,width=9cm, angle=0}
 \caption{Equivalent widths for the \ion{N}{ii} 3995\AA\ line (upper
   panel), \ion{C}{ii} doublet at 4267\AA\ (middle panel) and
   the \ion{Mg}{ii} doublet at 4481\AA\ (lower panel) plotted
   against effective temperature. The targets with \NtoH$\geq$7.4 are
   shown in red in the lower two panels.}
 \label{f_cii_mgii}
 \end{figure}

B-type stars have a rich \ion{O}{ii} spectrum \citep[see, for example,][]{sim06}. However, the very small changes predicted for the oxygen abundance would be masked by random errors in our estimates that, from the simulations of \citet{hun07}, are $\sim$0.2--0.3\,dex for oxygen. The depletions predicted for the carbon abundance could be discernable in the most nitrogen enhanced targets but this would depend on their initial mass and rotational velocity, the latter being unknown. The only \ion{C}{ii}\ feature observed in our spectra is the doublet at 4267\,\AA\ and we  have analysed its equivalent widths to deduce carbon abundances. \footnote{The \ion{C}{ii} doublet at 4267\AA\ can also be affected by nebular di-electronic emission, which can make reliable sky subtraction difficult especially for fibre spectroscopy \citep{eva11, mce14}. However inspection of the spectroscopy from the sky fibres showed no evidence for significant emission.} For the whole sample, a mean abundance of \CtoH=7.17$\pm$0.20\,dex was obtained, whilst for the targets with \NtoH>7.4 dex, the mean was \CtoH=7.17$\pm$0.18\,dex. Hence there is no evidence for a carbon depletion in the latter, although these results are not inconsistent with the predictions of \citet{bro11a}. In Fig.~\ref{f_cii_mgii} we show equivalent width estimates for the \ion{C}{ii} and \ion{Mg}{ii} doublets, which have a similar strength and dependence on effective temperature. These plots show scatter due to both observational uncertainties and the variations in gravity and microturbulent velocity of the targets. The stars with the largest nitrogen abundance estimates, \NtoH$>$7.4\,dex, have \ion{Mg}{ii} equivalent widths consistent with the other targets\footnote{ It was not possible to estimate \ion{Mg}{ii} and \ion{C}{ii} equivalent widths for the target \#1125, and a \ion{Mg}{ii} equivalent width for \#0023. All these targets have high effective temperatures, where these spectral lines are intrinsically weak.}, as would be expected for these relatively unevolved stars. The \ion{C}{ii} equivalent widths also show no evidence for large carbon depletions consistent with the mean abundances discussed above. The upper panel of Fig.~\ref{f_cii_mgii} shows the equivalent widths of the \ion{N}{ii} 3995\AA\ line as a function of effective temperature for our targets. In contrast to the behaviour of the \ion{C}{ii} and \ion{Mg}{ii} lines, there is no strong trend with effective temperature, indicating that other factors must influence its intensity.

 A helium to hydrogen abundance ratio by number of 0.1 was assumed in our calculations of theoretical spectra (see Sect.\ \ref{s_atm}). The SMC evolutionary models of \citet{bro11a} appropriate to B-type stars indicate that even for a nitrogen enhancement of 1.3 dex (larger than that seen in our sample), the change in the helium abundance is $\leq$0.05 dex. Similar small changes were predicted for the nitrogen-enriched targets in Paper I and implies that our assumption of a normal helium abundance is unlikely to be a significant source of error.
 
 As discussed in Sect.\ \ref{s_atm}, some targets appeared to have silicon abundances below the adopted baseline for the SMC. A possible explanation would be that they are unidentified binaries, where the \ion{Si}{iii} line strengths have been reduced by the continuum flux from an unseen secondary. \citet{gar17} analysed the VFTS spectroscopy of B-type SB1 targets and estimated nitrogen abundances. They found that none of the binaries showed the significant nitrogen enhancements found in some of the corresponding apparently single B-type sample in Paper I. For our nine apparently silicon weak targets, six have nitrogen enhancements of less than 0.4 dex,  whilst the upper limits of the other three targets imply enhancements of less than 0.6-0.7 dex. This would be consistent with some or all of these targets being undetected binaries.
 
 As well as estimating the stellar masses and ages (see Sect.\ \ref{s_M_A}), {\sc bonnsai} also predicts current nitrogen abundances. For all our targets, these corresponded to enhancements of 0.00-0.02\,dex, apart from \#0057 for which an enrichment of 0.07\,dex was predicted. Additionally, the predicted initial (\vi) and current (\ve) rotational velocities were small. For the cohort with \vsini$\leq$40\kms, these were all less than or equal to 40\,\kms, whilst for the cohort with 40\,$<$\,\vsini$\,\leq$\,80 \kms, they were generally in the range 50-90\,\kms. Two targets yielded larger estimates, viz. \#0057: \vi$\sim$110\,\kms, \ve$\sim$90\,\kms\ and \#1206: \vi$\sim$120\,\kms, \ve$\sim$120\,\kms. The small nitrogen enhancements (and correspondingly low rotational velocities) predicted by {\sc bonnsai} are consistent with the low numbers of nitrogen enhanced targets found in our simulations in Sect.\ \ref{d_sim}. 
 
 \subsection{Nitrogen enhanced targets} \label{d_N_en}
 
 In Paper I, the characteristics of the nitrogen-enhanced stars in 30 Doradus were compared to those of the rest of the sample. The former had on average higher effective temperatures, masses and luminosities and lower ages than the other stars. However these differences were relatively small and normally not significant at the 5\% statistical level. Here we undertake a similar investigation for our current sample adopting a threshold of \NtoH$>$6.8\,dex. We find similar trends to those in Paper I with the differences in the medians being 1070\,K for \teff,  0.24\,dex for logarithmic luminosity, 1.7\,\Msun\ for mass and $-$2.3\,Myr for age, with similar differences being found for the averages. Student t-tests and Mann-Whitney U-tests \citep{mod11} similar to those employed in Paper I are summarized in Table \ref{t_stat} and imply that the luminosity differences are significant at the 1\% level. Additionally the Student t-test find the mass differences significant at the 1\% level.
 
\begin{table}
\begin{center}
  \caption{Probabilities ($P$) that the estimated physical parameters
    of the nitrogen-enriched samples come from the same parent
    population as the other targets. The last two sets of
    probabilities include targets with a {\em possible} nitrogen
    enhancement of more than 0.3\,dex.}\label{t_stat}
\begin{tabular}{llrrrrr}
\hline\hline
\vsini &  Test & $P_{T{\rm eff}}$ & $P_L$~ & $P_M$~ & $P_{\rm Age}$\\
(\kms) &       & (\%) & (\%) & (\%) & (\%) \\
 			\hline
0-80  & Mann-Whitney    &  28.5 &  0.1 & 14.4 & 27.6 \\
      & Student t       &  15.8 & <0.1 &  0.5 & 14.8 \\
\\
0-40  & Mann-Whitney    &   9.5 & <0.1 &  0.5 & 12.9 \\
      & Student t       &   4.8 & <0.1 &  0.2 &  6.6 \\
\\
40-80 & Mann-Whitney    &  74.1 & 16.5 & 89.5 & 94.4 \\
      & Student t       &  36.4 &  2.2 & 15.2 & 45.1 \\
             \hline
0-40  & Mann-Whitney    &   7.0 &  0.9 &  1.2 & 10.5 \\
      & Student t       &   5.6 &  0.5 &  0.7 & 16.5 \\
\\
40-80 & Mann-Whitney    &  95.2 & 77.2 & 87.3 & 80.3 \\
      & Student t       &  48.1 & 23.6 & 23.6 & 44.3 \\
\hline
\end{tabular}
\end{center}
\end{table}

In Sect.\ \ref{d_sim}, we find that only the number of nitrogen-enriched targets with \vsini\,$\leq$\,40 \kms\ is inconsistent with evolutionary models incorporating rotational mixing. We have therefore repeated the above statistical test using subsets of nitrogen enhanced targets with \vsini\,$\leq$\,40 \kms\ and with 40\,$<$\,\vsini$\,\leq$\,80 \kms. These are again summarised in Table \ref{t_stat} and imply that the former (i.e.\ \vsini\,$\leq$\,40 \kms) have different masses and luminosities significant at a 1\% confidence level to the rest of the sample. By contrast in general the probabilities for the latter (40\,$<$\,\vsini$\,\leq$\,80 \kms) are consistent with the two samples being drawn from the same parent populations.
 
These statistical tests should be treated with some caution as there is the possibility of biases. For example, the targets with larger luminosities normally have better quality spectroscopy. This could lead to nitrogen enhancements being preferentially identified in such targets. To test this potential bias, we have arbitrarily assumed that all targets with a nitrogen abundance upper limit, \NtoH$>$6.8 dex, are nitrogen enhanced and have repeated the statistical tests. For the cohort with \vsini\,$\leq$\,40 \kms, there were 9 such targets and the statistical tests (see Table \ref{t_stat}) imply that the differences for the masses and luminosities remain significant normally at the 1\% level. For the 40\,$<$\,\vsini$\,\leq$\,80 \kms\ cohort, there were 8 additional targets. Now the probability for the Student t-test of the luminosities has been significantly increased and all percentages are now greater than 20\%. Hence we conclude that this bias is probably unimportant for the cohort with \vsini\,$\leq$\,40 \kms\ but would caution that other unidentified biases could be present.

The radial distribution of our nitrogen-enhanced targets is very similar to that of the rest of the sample with differences in their medians and averages of 0\farcm04 and $-$0\farcm85 respectively. This is consistent with the conclusion in Paper I that the distribution between field and cluster stars were similar for the nitrogen enriched and normal samples. 

The dynamics of our samples can also be investigated using our radial velocity estimates and the {\em Gaia} DR2 proper motions \citep{gaia1, gaia2} summarised in Table~\ref{t_Targets}. For our targets with nitrogen abundance estimates, we find mean proper motions of $\mu_{\rm RA}$\,$=$\,0.87$\pm$0.24\,\pmg\ and $\mu_{\rm Dec}$\,$=$\,$-$1.24$\pm$0.15\,\pmg. These are similar to the median values of $\mu_{\rm RA}$\,$=$\,0.80\,\pmg\ and $\mu_{\rm Dec}$\,$=$\,$-$1.22\,\pmg\ estimated in Paper II for targets with $G$\,$\leq$\,16, which lay within a 10\arcmin\ radius centred on NGC\,346. Additionally the mean radial velocity for our whole sample is \vr$=$161$\pm$19\,\kms.
 
\begin{figure}
	\epsfig{file=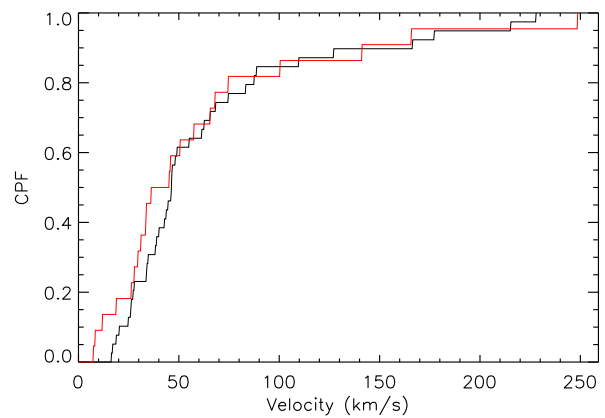,width=9cm, angle=0}\\
	\caption{Cumulative probability functions for the space velocity relative to the mean velocity for the nitrogen-enhanced stars (red line) and the rest of the sample (black line).}
	\label{f_cpf_vel}
\end{figure}

The corresponding means for the 22 nitrogen-enhanced targets are very similar with values: $\mu_{\rm RA}$\,$=$\,0.87$\pm$0.26\,\pmg, $\mu_{\rm Dec}$\,$=$\,$-$1.21$\pm$0.12\,\pmg\, and \vr$=$163$\pm$20\,\kms. In turn this implies that, unsurprisingly, there are no significant {\em systematic} kinematic differences between the two samples. In Fig.\ \ref{f_cpf_vel}. the cumulative probability function for the space velocity with respect to the mean velocity are shown for both the nitrogen-enhanced targets and the rest of the sample. These again appear very similar and this is confirmed by a Kolmogorov-Smirnov test \citep{fas87}  that returns a high probability of 59\% that they were taken from the same parent population.

One nitrogen enhanced target (\#0026$\equiv$MPG\,12) appears to have an anomalously large radial velocity and additionally its RA proper motion is amongst the smallest of our dataset. \citet{wal00} previously commented on both the strength of its nitrogen spectrum and  its discrepant radial velocity -- their estimate of 226\,\kms\ agrees well with that obtained here. Additionally \citet{bou13} have obtained very similar atmospheric parameters (\teff$\sim$31\,000K, \logg$\sim$3.65\,dex) and nitrogen abundance (\NtoH=7.70$\pm$0.21\,dex). This star could be a runaway and have an evolutionary history different to those of our other nitrogen enhanced stars.
 
To summarise, the nitrogen-enhanced stars appear to have similar spatial and kinematic properties to the rest of the sample. They have higher median effective temperatures, masses and luminosities and lower median ages in agreement with the results for a similar sample of LMC targets discussed in Paper~I. For the cohort with \vsini\,$\leq$\,40 \kms, statistical tests imply that these differences are significant for the masses and luminosities, although there remains the possibility of unidentified biases.

\begin{table*}

\caption{Observed and predicted numbers of nitrogen-rich targets with enhancements of greater than 0.3 dex (\NtoH$\geq$6.8 dex), 0.6 dex (\NtoH$\geq$7.1 dex) and 0.9 dex(\NtoH$\geq$7.4 dex). The Monte-Carlo simulations are designated as `MC'. Simulations are provided for two ranges of projected rotational velocity, \vsini, viz. 0-40 \kms\ and 40-80 \kms.}\label{t_N_sim}

\begin{center}
\begin{tabular}{lrrrrrrrrrr}
\hline\hline
Sample  &  $N$ & \multicolumn{2}{c}{\NtoH$\geq$6.8} &  \multicolumn{2}{c}{\NtoH$\geq$7.1}  &\multicolumn{2}{c}{\NtoH$\geq$7.4} \\
\vsini\ (\kms)&     & 0-40 & 40-80 & 0-40 & 40-80 & 0-40 & 40-80\\   \hline  
&&&&&\\ 
Observed      & 211 & 14            & 8             & 7             & 7             & 4             & 4             \\
&&&&&\\
MC            & 211 & 2.7$\pm$1.6 & 8.0$\pm$2.8 & 1.7$\pm$1.3 & 5.0$\pm$2.2 & 0.6$\pm$0.8 & 1.6$\pm$1.3 \\ 
MC Prob  & 211 & 0.00\%       & 59.49\%           & 0.18\%          & 23.48\%           & 0.29\%          & 8.43\%           \\ 
 \hline
\end{tabular}
\end{center}
\end{table*}

\subsection{Numbers of nitrogen enriched targets expected from rotational mixing} \label{d_sim}

As discussed above, it is possible that the targets with the largest nitrogen enhancements could have evolved as single stars undergoing rotational mixing. Following the methodology of Paper~I, we have estimated the number of rotationally-mixed stars that we might expect to find in our sample. Only a brief outline of the methodology will be provided as a detailed description is given in Paper~I.

We have considered three thresholds for nitrogen enrichment, viz. 0.3 dex (corresponding to a nitrogen abundance, \NtoH$>$6.8 dex), 0.6 dex (\NtoH$>$7.1 dex) and  0.9 dex (\NtoH$>$7.4 dex). The first was chosen so that any enhancement would be comparable to the estimated errors in the nitrogen abundances, whilst the latter two were chosen to identify more highly-enriched targets. The number of stars fulfilling these criteria are summarised in Table \ref{t_N_sim}.

From the SMC evolutionary grid of \citet{bro11a}, the stellar rotational velocities that would provide such enrichments can be estimated. As in Paper I, the core hydrogen burning phase, where the surface gravity, \logg $>$ 3.2 dex, was considered so as to be consistent with the gravities estimated for our samples (see Tables \ref{t_Atm1} and \ref{t_Atm2}). In Table \ref{t_v_e}, the {\em minimum} rotational velocities, \vl, that a star would have during this evolutionary phase are listed for initial masses of 5-16\Msun, compatible with the mass estimates in Tables \ref{t_Atm1} and \ref{t_Atm2}. For our simulations, we have taken a conservative approach and adopted velocities (see Table \ref{t_v_e}) towards the lower ends of these estimates. 

As discussed in Paper I, the number of targets (assuming a random distribution of axes of inclination) that have both \vsini\,$\leq v'$ and a nitrogen enhancement, \NtoH$\geq$\NtoHL\  can be written as:

\noindent 
\begin{equation}\label{e_1}
\begin{split}
N(v_e\sin i \leq v'; \epsilon \geq \epsilon_\mathrm{0})=N\int^{\infty}_{v_\mathrm{0}} f(v_e)\left [1- \cos i_0 \right ]dv_e
\end{split}
\end{equation}

\noindent where $N$\ is the total number of targets and

\noindent 
\begin{equation}\label{e_2}
i_0= \sin^{-1}\left(\frac{v'}{v_e}\right )
\end{equation}

\noindent and $f(v_e)$ is the normalised distribution of rotation velocities and has been taken from Paper II.

Equation \ref{e_1} can be integrated  numerically for different values of $v'$\ and $\epsilon_\mathrm{0}$\ by adopting the appropriate value of \vl\ from Table~\ref{t_v_e}. An alternative methodology also used in Paper I was to adopt a Monte-Carlo approach, where samples of $N$\ targets were randomly assigned both rotational velocities (using the rotational velocity distribution, $f(v_e)$\ discussed above) and  angles of inclination. The number of targets with different nitrogen enrichments and in different projected rotational bins are then found using the rotational velocity limits, \vl,  in Table \ref{t_v_e}. 

\begin{table}
\begin{center}
\caption{Estimates of the minimum rotational velocities, \vl, predicted by the SMC evolutionary models of \citet{bro11a} required to achieve different nitrogen enhancements whilst maintaining a gravity, \logg$>$ 3.2 dex. Enhancements of greater than 0.3 dex (\NtoH$\geq$6.8 dex), 0.6 dex (\NtoH$\geq$7.1 dex)  and  0.9 dex (\NtoH$\geq$7.4 dex) are considered. Initial masses (\Mi) are in units of the solar mass, \Msun. The values characterised as `Adopted' were used in the simulations discussed in Sect.~\ref{d_sim}.}\label{t_v_e}
\begin{tabular}{lccc}
\hline\hline
\Mi/\Msun &  \vl(6.8\,dex) &  \vl(7.1\,dex) &  \vl(7.4\,dex)\\
                   &\kms&\kms&\kms\\   \hline  
\,\,\,5 & 111 & 184 & 314 \\
\,\,\,8 & 110 & 173 & 287 \\
12      & 113 & 166 & 272 \\
16      & 118 & 157 & 261 \\ 
Adopted & 115 & 160 & 270 \\
\hline
\end{tabular}
\end{center}
\end{table}

\subsection{Comparison of simulations with numbers of observed nitrogen enriched targets} \label{d_obs}

In Paper I, it was found that the numerical integrations and the Monte Carlo simulations led to results that were indistinguishable and similar agreement has been found here. Hence, in Table \ref{t_N_sim}, only the Monte-Carlo simulations are summarised as they also provide estimates of the standard errors. The latter are consistent with Poisson statistics as would be expected given the low probability of observing stars at small angles of inclination. These simulations appear to significantly underestimate the number of nitrogen enriched targets with \vsini\,$\leq$\,40 \kms, whilst being consistent with cohort having 40\,$<$\,\vsini$\,\leq$\,80 \kms, similar to the results in Paper I for the LMC samples.

The Monte-Carlo simulations predict the probabilities of observing a given number of nitrogen-enriched stars. When the {\em observed number}, $n$ was greater than predicted, we have estimated the probability that the number observed would be greater than or equal to $n$. Conversely, when the observed number, $n$ was less than predicted, we have calculated the probability that the number observed would be less than or equal to $n$. These probabilities are also summarised in Table \ref{t_N_sim} and again very similar probabilities would have been found from simply adopting Poisson statistics. For the targets with \vsini\,$\leq$\,40 \kms, all the simulations imply that too many nitrogen-enriched stars are observed at a high degree of probability ($>$99\%). This is especially the case for a nitrogen enrichment, \NtoH$\geq$6.8\,dex, where the simulation (using a million iterations) returned no cases where 14 (or more) targets would be observed; adopting Poisson statistics would have given a probability of $1.0\times 10^{-4}$\%. By contrast for the samples with  40\,$<$\,\vsini$\,\leq$\,80~\kms, the agreement between the observations and the simulations is good. Only for the most nitrogen-enhanced stars are the number of observed targets noticeably greater than predicted and even then this is not significant at a 5\% level. One of the observed targets (\#0026) may be a runaway with a different evolutionary history (see Sect.\ \ref{d_N_en}); excluding this target would increase the probability for the cohort with \NtoH$\geq$7.4\,dex and 40\,$<$\,\vsini$\,\leq$\,80~\kms\ to 23\%.

For the VFTS Monte-Carlo simulations, a total sample of $N=211$ was adopted. This corresponds to all the apparently single targets in Paper II that have an estimated gravity of \logg$>$3.2\,dex, and was chosen so as to be consistent with the cut-off adopted in the simulations. The predicted number of targets are 35.6$\pm$5.5 and 27.6$\pm$4.9\ with \vsini\,$\leq$\,40 \kms\ and  40\,$<$\,\vsini$\,\leq$\,80 \kms\ respectively. These are in good agreement with the 37 and 26 targets in the original sample in Table~\ref{t_Targets}. The small differences probably reflect the uncertainties in the adopted rotational distribution function, $f(v_e)$, which have been discussed in Paper II.

The comparison of observed and simulated numbers of nitrogen-enhanced targets is relatively insensitive to the choice of the nitrogen baseline abundance (see Sect.~\ref{d_N}). For example increasing this by 0.1\,dex to \NtoH=6.6\,dex, the number of targets with vsini\,$\leq$\,40 \kms\  and nitrogen enhancements of at least 0.3\,dex (\NtoH$>$6.9\,dex),  0.6\,dex (\NtoH$>$7.2\,dex) and 0.9\,dex (\NtoH$>$7.2\,dex) would be reduced to 11, 6 and 3 respectively. In turn adopting the current simulations, the probabilities would be increased to 0.01\%, 0.81\% and 2.03\% respectively. These simulations are based on the \vl\ estimates listed in Table \ref{t_v_e}, which are taken from models assuming a baseline abundance of \NtoH=6.5\,dex. In Paper I, larger \vl\ estimates were obtained for the LMC and a baseline nitrogen abundance, \NtoH=6.9\,dex. Hence increasing our baseline nitrogen SMC abundance would probably increase the \vl\ estimates, thereby decreasing the number of targets in the simulations. In turn this would increase the discrepancy for the cohort with \vsini\,$\leq$\,40 \kms.

In summary, the simulations presented in Table \ref{t_N_sim} are generally consistent with the nitrogen enriched targets with 40\,$<$\,\vsini$\,\leq$\,80 \kms\ having large rotational velocities (and small angles of incidence) leading to significant rotational mixing. By contrast, there would appear to be too many nitrogen-enriched targets with \vsini\,$\leq$\,40 \kms\ to be accounted for by this mechanism. These findings are similar to those found in Paper I for the LMC samples.

\subsection{Frequency of nitrogen enriched, slowly rotating  B-type stars} \label{d_freq}

Estimation of the frequency of nitrogen-enriched targets that cannot be explained by rotational mixing in a single star is not straightforward. Firstly, it is necessary to set a minimum threshold on the observed nitrogen enrichment. This requires a balance between identifying stars that are indeed nitrogen rich against setting such a high threshold that this frequency is seriously underestimated. The latter is illustrated in Table \ref{t_N_sim}, where the observed excess of nitrogen-enriched stars in the cohort with \vsini\,$\leq$\,40 \kms\ increases as the threshold is decreased. Given that we believe that our nitrogen abundances have a typical uncertainty of $\pm$0.2--0.3\,dex, we have decided to set the threshold at an enrichment of at least 0.3\,dex, i.e.\ \NtoH$\geq$6.8 dex. Secondly, the simulations presented in Sect.\ \ref{d_obs} imply that there is an excess of nitrogen-enriched targets for targets \vsini\,$\leq$\,40 \kms, whilst there is no significant evidence for such objects in the cohort with 40\,$<$\,\vsini$\,\leq$\,80 \kms. This implies that these objects have {\em rotational velocities}, $v_e\leq 40$\kms\ and this will be assumed in the discussion below. 

Following the methodology of Paper I, we will consider two measures for the frequency of nitrogen-enhanced objects. Firstly the percentage, $P_{40}$, of stars {\em within the cohort with $v_e\leq$40 \kms} that have enhanced nitrogen abundances that do not appear to be due to rotational mixing and secondly the percentage, $P_{\mathrm{T}}$, of such stars within the total population of single stars. The adopted rotational velocity distribution, $f(v_e)$\ taken from Paper II implies that 11.5\% of our 211 apparently single targets will have \ve\,$\leq$\,40 \kms. This translates into a number of targets, N$_{40}\sim24$ and corresponding to 65\% of the targets with \vsini$\leq 40$\kms; a similar ratio to the 67\% found for the VFTS targets in Paper I.

The excess of targets ($E_{40}$) with nitrogen abundances, \NtoH$\geq$6.8 dex and $v_e\leq 40$\kms\ can be deduced from the results in Table \ref{t_N_sim} and these estimates are summarized in Table \ref{t_N_exc}. Using the number of targets discussed above, the percentages of such targets can then be estimated and are again summarised in Table \ref{t_N_exc}. It is possible that the number of nitrogen-enriched targets have been underestimated as in some cases it was only possible to set upper limits on this abundance. For the cohort with $v_e\leq 40$\kms, 9 targets have limits consistent with a nitrogen abundance, \NtoH$\geq$6.8\,dex. Including all these targets would provide upper limits for these percentages and these have also been summarised in the lower two rows of Table \ref{t_N_exc}. Using the results discussed in Paper I, it is possible to deduce similar percentages for the two LMC samples from the VFTS and the FSMS. These are also summarised in Table \ref{t_N_exc} with no upper limits being given for the FSMS as nitrogen abundances could be estimated for all the targets.

\begin{table}
\begin{center}
\caption{Estimates of the excess number ($E_{40}$) of nitrogen-rich targets with enhancements of greater than 0.3 dex that are not due to rotational mixing in single stars. Also listed are the total number of apparently single B-type stars ($N$) and the inferred number ($N_{40}$) of these targets that have \ve$\leq$40\,\kms . These have been used to estimate the corresponding percentages compared with the total number of apparently single B-type stars ($P_{\mathrm{T}}$) and with the number of targets that have \ve$\leq$40\,\kms ($P_{40}$)}\label{t_N_exc}
\begin{tabular}{lrrr}
			\hline\hline
			Sample           & NGC346     & VFTS        & FSMS  \\
			Galaxy           & SMC        & LMC         & LMC   \\
\hline
\\
			$E_{40}$         & 11.3        & 5.0         & 5.1   \\
			$N$              & 211        & 220         & 103   \\
			$N_{40}$         & 24         & 21          & 12    \\
			\\
			$P_{\mathrm{T}}$(\%) & 5.4        & 2.3         & 4.9   \\
			$P_{40}$(\%)         & 47         & 24          & 42    \\
			\\
			$P_{\mathrm{T}}$(\%) & $\leq9.6$  & $\leq5.5$   & -     \\
			$P_{40}$(\%)         & $\leq85$ & $\leq57$  & -     \\			
\hline
		\end{tabular}
	\end{center}
\end{table}

For all three datasets, the excess of the nitrogen-enhanced stars are a small percentage of the total B-type population. However they constitute a significant percentage of targets with \ve$\leq40$\,\kms, $P_{40}\sim24-47$\%. Indeed, allowing for targets that have only upper limits for their nitrogen abundance implies that this percentage could be $>$50\% for the NGC\,346 and VFTS datasets.

Additionally, these percentages may have been underestimated for several reasons. Firstly, most of our targets have larger gravities than the limit of \logg$>$3.2\,dex adopted for the rotational-velocity estimates in Table \ref{t_v_e}; indeed the median gravity of our sample is 4.0\,dex. As discussed in Paper I, this may lead to an overestimation of the number of nitrogen-enriched targets in our simulations. However given that these estimates are considerably smaller than those observed, this is unlikely to be a major source of error. For example, it could increase the $P_{40}$\ for the NGC\,346 sample to at most 58\%. Secondly, by setting a threshold at a minimum enrichment of 0.3 dex, modestly nitrogen-enriched targets will have been excluded. For example as discussed in Paper I, decreasing the threshold to a minimum enrichment of 0.2\,dex would lead to estimates of $P_{40}$ of  $\leq$76\% and $\sim$71\% for the VFTS and FSMS samples respectively.

Although for all three samples, our targets do not show significant radial velocity variations, they will contain undetected binary systems. \citet{dun15} found an {\em observed} binary fraction for the non-supergiant VFTS B-type stars of 25$\pm$2\%. However from a Monte-Carlo simulation, they estimated that $\sim$45\% of the remaining apparently single stars would be binaries. Similarly large percentages of undetected binaries would be expected in the other samples given the limited time cadence of their observations \citep[see] [and Paper II for details]{eva06}. For the VFTS, \citet{gar17} has analysed the B-type targets identified as SB1 systems. They found that these targets did not exhibit significant nitrogen enhancements unlike the analysis of the apparently single VFTS targets discussed in Paper I. If this lack of nitrogen enrichment in binaries was also present in the undetected binaries in our datasets, it would lead to the nitrogen-enhanced targets being concentrated in the subsets of actual single stars. This would in turn significantly increase the $P_{40}$\ estimates. For example, adopting the undetected binary estimate of \citet{dun15} for the VFTS would increase its  $P_{40}$ to $\sim$44\%, whilst the upper limit would now be 100\%, with similar results being found for the other datasets. Indeed if the nitrogen enriched targets were solely (or predominantly) single stars and the two other possible biases discussed above were also present, it is possible that $P_{40}\sim$100\% for all three datasets.

In summary for all three samples, we find $P_{\mathrm{T}}$$\sim$2--6\% and $P_{40}$$\sim$24--47\%. Including targets with only upper limits for their nitrogen abundance estimates would lead  $P_{\mathrm{T}}\la$9\% and $P_{40}\la70$\% for the NGC\,346 and VFTS samples. Given the methodology, these percentages should probably be considered as lower limits. Indeed the adoption of conservative \vl\ estimates, the exclusion of targets with modest nitrogen enhancements and the presence of undetected binaries in our samples imply that it is possible that all the single targets with \ve$\leq$40\,\kms\ have nitrogen enhancements that are inconsistent with rotational mixing, i.e. $P_{40}\sim$100\% for all three datasets.

\subsection{Origin of the nitrogen enriched targets} \label{d_origin}

In Paper I, several explanations were considered to explain the excess of nitrogen-enriched stars with \vsini\,$\leq$\,40 \kms\ -- hereafter referred to as simply `nitrogen enhanced'. These included the physical assumptions in standard single-star evolutionary models, binarity and magnetic fields, together with combinations of these. For the first, the efficiency of rotational mixing and the role of mass loss were considered. However it was unclear how these processes could lead to such different nitrogen abundances in targets covering a relatively small range of spectral-types and this remains the case for our current sample. Another possible explanation was that they were {\em extant} binary systems. However, the analysis of \citet{gar17} of the narrow-lined B-type SB1 systems from the VFTS showed that none of the primaries had significant nitrogen enrichment. They concluded that these systems were in a pre-interaction epoch and evolving as single stars with low rotational velocities and small amounts of rotational mixing \citep{deM09}. Hence neither of these explanations will be considered further but more details can be found in Paper I.

The two most promising explanations for the nitrogen-enriched targets that were identified in Paper I were post-interaction binaries or magnetic fields in single stars. Post-interaction binaries will probably be classified as single stars on the basis of a lack of radial velocity variations \citep{deM14}. This could be due to either a stellar merger or alternatively following mass transfer, the mass gainer would dominate the light from the system (precluding an SB2 classification), whilst typically only having radial velocity variations of less than 20\,\kms. However as discussed in Paper I, most of the evolutionary pathways for such systems  \citep{lan08, bro11b} would be expected to produce large rotational velocities. Indeed they could explain the presence of a high-velocity tail (\vsini\,$> 300$\,\kms) found in the rotational velocity distribution of the VFTS `single' O-type star sample \citep{ram13, deM13}.

However magnetic fields could also play a role in mergers of pre-main or main sequence stars \citep{pod92, sta10, lan12, kor12, sch16}  as this could provide an efficient spin down mechanism. Nitrogen enrichments might then be due to either the merger and its aftermath or rotational mixing of the initially rapidly rotating product. Products of such mergers may represent $\sim$10\% of the O-type field population \citep{deM14} and if the fraction was similar in the B-type population, this would be consistent with the percentages of nitrogen-enriched stars summarised in Table \ref{t_N_exc}. Additionally, as discussed in Paper I, the products of such mergers might be expected to be more massive and appear younger than other single B-type stars, compatible with the results found here and in Paper I. 

Very recently \citet{sch19} have undertaken three-dimensional magnetohydrodynamical simulations of the coalescence of two massive stars and compared their predicted parameters to those for the nitrogen enhanced star, $\tau$~Sco \citep{mok05, sim06, nie14}. Good agreement is found for both the atmospheric parameters, small rotational velocity and large magnetic field. Several of our targets (e.g.\ \#0103, \#1125) and those discussed in Paper I (VFTS095, 214, 692) have similar atmospheric parameters to those of $\tau$~Sco and could be Magellanic Cloud analogs. 

Magnetic fields in single stars have been previously suggested as an explanation for nitrogen enriched, slowly rotating {\em Galactic} early-B stars \citep{mor08,mor12,prz11}, although \citet{aer14a} found no significant correlation between nitrogen abundances and magnetic field strengths in a statistical analysis of 68 Galactic O-type stars. \citet{mey11} calculated evolutionary models including magnetic fields with both differential and solid-body rotation during the main sequence phase. Both sets of models produced slowly rotating stars but only the former had significant surface nitrogen enhancements. 

Recently, \cite{kes19} have calculated further models which included the magnetic field evolution and mass-loss magnetic quenching, which were not considered by \citet{mey11}. Again they find that the models evolve to low rotational velocities. However although the model with differential rotation shows the greater nitrogen enhancement,  a significant enhancement is also found in the model with solid body rotation. \cite{kes19} comment that this behaviour would explain the `Group 2' stars introduced by \citet{hun08b}. These stars had low projected rotational velocities and significant nitrogen abundance enhancements and are effectively the FSMS sample discussed in Paper I and here (see Table \ref{t_N_exc}).

This explanation must be treated with some caution as all their models adopt a Galactic nitrogen abundance, which is between 0.6-1.0 dex higher than those in the Magellanic Clouds. This could lead to nitrogen enhancements being considerably different in these environments. As an exemplar, we consider the maximum main-sequence nitrogen enhancement for (non-magnetic) stars with initial rotational velocities of approximately 280\,\kms\ and an initial stellar mass of 15\Msun. We use the models of \cite{bro11a} that adopt realistic initial Galactic and Magellanic Cloud chemical compositions. Nitrogen enhancements of 0.41\,dex (Galactic), 0.75\,dex (LMC) and 0.97\,dex (SMC) are found, with their significant range illustrating the difficulty of using models with an inappropriate initial composition in any quantitative comparison with observation.
 
However a qualitative comparison can be made of the models of \cite{kes19} with our SMC results and those for the LMC from Paper I. The Galactic evolutionary models including rotation and magnetic field predict a nitrogen enhancement at the terminal age main sequence of approximately 0.6-0.8 dex. This compares with a maximum observational enhancement of approximately 1.3\,dex (SMC, this paper) and 1.0\,dex (LMC, Paper I). These would appear to be consistent given the different baseline chemical compositions discussed above. Additionally the evolutionary models imply that most of the nitrogen enhancement occurs relatively early in the main sequence lifetime. For example when the core hydrogen abundance has dropped by 50\% (corresponding to a surface gravity of approximately 3.9\,dex), the predicted nitrogen enhancement is 0.5-0.8\,dex. This is again consistent with some of the observed nitrogen-enhanced targets in both the SMC (this paper) and the LMC (Paper I) having relatively large gravities. Finally, the evolutionary models imply very low rotational velocities ($<$3\,\kms\ at TAMS) and this could provide a strong observational constraint. Unfortunately the current observational lower limit on the projected rotational velocity is approximately 40\,\kms\ and pushing this limit to lower values would require significant observational resources.

In Paper I, the percentage of {\em Galactic} B-type stars that were magnetic was discussed. Estimates were available from the the MiMes survey \citep{wad14, wad16} and the BOB survey \citep{mor14, mor15} and typically ranged from 5-8\% \citep{wad14, fos15a,sch17} with a similar frequency for O-type stars \citep{gru17}. Assuming that the frequencies are similar in the lower metallicity environments of the Magellanic Clouds, these are consistent with the frequency of nitrogen enhanced targets ($P_{\mathrm{T}}$) summarised in Table \ref{t_N_exc}.

\section{Concluding remarks}

An excess of B-type stars with small projected rotational velocities and significant nitrogen enhancements was first identified by \citet{hun08b,hun09a} and designated `Group 2' stars. The nature of these targets was subsequently discussed by \citet{bro11b}, \citet{mae14}, and \citet{aer14a}, whilst similar O-type stars were identified by \citet{riv12}, and \citet{gri17}. In Paper~I, the targets identified by \citet{hun08b,hun09a} were re-considered and computer simulations were undertaken to investigate whether they could be rapidly rotating single stars observed at low angles of inclination. Similar analyses have been undertaken in Paper I for the VFTS LMC targets and here for SMC targets. Combined these datasets contained approximately 1000 apparently single B-type stars, of which approximately 200 had \vsini\ estimates of less than 80\,\kms. All three analyses implied:
\begin{enumerate}
\item  The {\em relative} numbers of nitrogen-enhanced targets in the cohorts with projected rotational velocities, \vsini\ of 0--40\,\kms\ and 40--80\,\kms\ are inconsistent with them all being rapidly-rotating stars observed at small angles of inclination.
\item  The numbers of nitrogen-enhanced targets in the cohort with projected rotational velocities of 40--80\,\kms\ are consistent with them being rapidly-rotating targets observed at relatively small angles of inclination.
\item The number of nitrogen-enriched targets with  \vsini\,$\leq$\,40\,\kms\ is larger than predicted. These differences are significant at a high level of probability for {\em all three datasets}.
\end{enumerate}
These results imply that the excess of nitrogen-enriched stars with \vsini\,$\leq$\,40\,\kms\ have rotational velocities, \ve$<$40\,\kms. As such, they make up a relatively small percentage of the total apparently single B-type stellar population but a significant percentage of the population with \ve$<$40\,\kms. These have been estimated to be $P_{\mathrm{T}}\sim 5$\% and  $P_{\mathrm{40}}\sim 40$\% respectively (see Table \ref{t_N_exc}). However, the nature of the simulations, the targets with only upper limits on their nitrogen abundances, the threshold used to define nitrogen-enriched stars and (possibly most importantly) the presence of unidentified binaries in our observational samples imply that these estimates should be considered lower limits. Indeed it is possible that all the B-type stars with \ve$<$40\,\kms\ are nitrogen enriched. 

Irrespective of the actual percentages there is now a considerable body of evidence that there are slowly rotating, hydrogen core burning B-type stars whose nitrogen abundances cannot be explained by rotational mixing. Similar stars may also exist in the corresponding O-type population.

Two attractive explanations for the properties of these stars incorporate the effects of magnetic field, being a stellar merger followed by magnetic breaking or the evolution of a single star with a large magnetic field. Both mechanisms would appear to be compatible with the observed frequency of nitrogen-enriched stars in the Magellanic Clouds, although the comparison is complicated by variations in the ambient chemical composition. The differences for all three datasets in the median effective temperatures, masses, luminosities and ages for the nitrogen-enriched stars compared with the rest of the sample would be consistent with the former mechanism. For the latter, a qualitative comparison with Galactic evolutionary models incorporating magnetic fields is encouraging in terms of the amount of nitrogen enrichment and its presence in stars near the zero-age main sequence. Of course it is possible that both mechanisms contribute to the producing slowly rotating nitrogen-enhanced B-type stars.

Further progress in understanding the evolutionary status of these Magellanic Cloud stars is likely to be challenging. Two observational approaches would be to identify the frequency of B-type stars in the Magellanic Clouds with magnetic fields or to better constrain the \vsini\ estimates of the nitrogen-enhanced stars; either of these approaches would be observationally very difficult. Evolutionary calculations of magnetic B-type stars with realistic chemical compositions (including variations in the CNO abundance ratios) for the Magellanic Clouds would be useful as they would allow a more robust comparison with observations. Given the large number of B-type stars considered in the current datasets, any improvement will be observationally demanding in terms of sample size and target selection. However, in principle, information on nitrogen enrichment as a function of apparent main-sequence lifetime would be very useful.

\begin{acknowledgements}
Based on observations at the European Southern Observatory Very Large Telescope. This work has made use of data from the European Space Agency (ESA) mission {\it Gaia} (\url{https://www.cosmos.esa.int/gaia}), processed by the {\it Gaia} Data Processing and Analysis Consortium (DPAC,
\url{https://www.cosmos.esa.int/web/gaia/dpac/consortium}). Funding for the DPAC has been provided by national institutions, in particular the institutions participating in the {\it Gaia} Multilateral Agreement. 
\end{acknowledgements}

\bibliography{36921.bib}

\end{document}